\documentclass[12pt,draftclsnofoot,onecolumn]{IEEEtran}

\usepackage{amsmath,amsfonts,amsthm,amssymb}
\usepackage[linesnumbered,ruled,vlined]{algorithm2e}
\usepackage{bbm}
\usepackage{graphicx}
\usepackage{cite}
\usepackage{color}
\usepackage{epstopdf}
\usepackage{float}
\usepackage{subfigure}
\usepackage{setspace}
\usepackage{stfloats}
\usepackage{url}
\usepackage{multirow}
\usepackage{gensymb}

\SetKwInput{KwInput}{Input}
\SetKwInput{KwOutput}{Output}
\DeclareMathOperator{\rank}{rank}
\DeclareMathOperator{\diag}{diag}
\DeclareMathOperator{\tr}{tr}

\def\BibTeX{{\rm B\kern-.05em{\sc i\kern-.025em b}\kern-.08em
    T\kern-.1667em\lower.7ex\hbox{E}\kern-.125emX}}

\begin{document}

\title{\begin{spacing}{1.3}Double-IRS Aided MIMO Communication under LoS Channels: Capacity Maximization and Scaling\end{spacing}}

\author{\begin{spacing}{1.3}Yitao~Han,~\IEEEmembership{Student~Member,~IEEE,}
	Shuowen~Zhang,~\IEEEmembership{Member,~IEEE,}\\
        Lingjie~Duan,~\IEEEmembership{Senior~Member,~IEEE,}
        and~Rui~Zhang,~\IEEEmembership{Fellow,~IEEE}\end{spacing}
\begin{spacing}{1.2}\thanks{Y.~Han and L.~Duan are with the Engineering Systems and Design Pillar, Singapore University of Technology and Design (e-mail: yitao\_han@mymail.sutd.edu.sg, lingjie\_duan@sutd.edu.sg). Y.~Han is also with the Department of Electrical and Computer Engineering, National University of Singapore.}
\thanks{S.~Zhang is with the Department of Electronic and Information Engineering, The Hong Kong Polytechnic University (e-mail: shuowen.zhang@polyu.edu.hk). S.~Zhang is the corresponding author.}
\thanks{R.~Zhang is with the Department of Electrical and Computer Engineering, National University of Singapore (e-mail: elezhang@nus.edu.sg).}\end{spacing}
}

\maketitle

\vspace{-6em}

\begin{abstract}
\vspace{-1em}
\begin{spacing}{1.45}
Intelligent reflecting surface (IRS) is a promising technology to extend the wireless signal coverage and support the high performance communication. By intelligently adjusting the reflection coefficients of a large number of passive reflecting elements, the IRS can modify the wireless propagation environment in favour of signal transmission. Different from most of the prior works which did not consider any cooperation between IRSs, in this work we propose and study a cooperative double-IRS aided multiple-input multiple-output (MIMO) communication system under the line-of-sight (LoS) propagation channels. We investigate the capacity maximization problem by jointly optimizing the transmit covariance matrix and the passive beamforming matrices of the two cooperative IRSs. Although the above problem is non-convex and difficult to solve, we transform and simplify the original problem by exploiting a tractable characterization of the LoS channels. Then we develop a novel low-complexity algorithm whose complexity is independent of the number of IRS elements. Moreover, we analyze the capacity scaling orders of the double-IRS aided MIMO system with respect to an asymptotically large number of IRS elements or transmit power, which significantly outperform those of the conventional single-IRS aided MIMO system, thanks to the cooperative passive beamforming gain brought by the double-reflection link and the spatial multiplexing gain harvested from the two single-reflection links. Extensive numerical results are provided to show that by exploiting the LoS channel properties, our proposed algorithm can achieve a desirable performance with low computational time. Also, our capacity scaling analysis is validated, and the double-IRS system is shown to achieve a much higher rate than its single-IRS counterpart as long as the number of IRS elements or the transmit power is not small.
\end{spacing}
\end{abstract}
\vspace{-1em}
\begin{IEEEkeywords}
\vspace{-1em}
\begin{spacing}{1.45}
Intelligent reflecting surface (IRS), multiple-input multiple-output (MIMO), double IRSs, alternating optimization, capacity scaling order.
\end{spacing}
\end{IEEEkeywords}

\section{Introduction}

The ever-growing demand for higher data rate, lower latency and enhanced reliability in wireless communications has driven both industry and academia to advance the communication technologies. Among them, massive multiple-input multiple-output (MIMO), millimeter wave (mmWave), and ultra-dense network (UDN) are the prominent candidates \cite{andrews20145G}. However, they incur high costs in energy consumption and hardware investment, by requiring a large number of antennas and the expensive radio frequency (RF) chains operating at high frequency bands. Moreover, the ever-increasing number of base stations (BSs) and access points (APs) can also generate high interference to each other and deteriorate the overall performance.

With the recent research advances in micro electromechanical systems (MEMS) and metamaterial \cite{cui2014coding}, it is now feasible to realize an amplitude change and phase shift to the incident signal in real time via a programmable surface, which enables an innovative wireless device and network component---intelligent reflecting surface (IRS) \cite{wu2020towards}. The IRS is a planar reconfigurable metasurface, which consists of a large number of passive reflecting elements and a smart controller. By adaptively configuring the reflection amplitude and phase of each element, the IRS can reshape the electromagnetic environment to fit specific needs, e.g., extending signal coverage, mitigating interference, and supporting massive device-to-device (D2D) communications \cite{zhang2020tutorial}. Different from the conventional AP or relay, the IRS only uses passive reflection and does not require high energy consumption or expensive hardware, which also gives it the potential to be densely deployed to significantly improve the network performance \cite{zhang2020tutorial}.

To fully gain the IRS's benefits, it is crucial to properly design its passive beamforming, which is a key focus of the IRS research (\!\cite{han2020cooperative,zhang2020capacity,yang2020ofdmwcl,yang2020ofdm,tao2021random,wu2019joint,huang2019ee,ye2020ser,guo2020sum,hou2020noma,ding2020noma,yu2020robust,pan2020multicell,cui2019secure,you2020double}). For example, \cite{wu2019joint} considered the joint optimization of BS's active beamforming and IRS's passive beamforming. \cite{zhang2020capacity} studied the capacity maximization of a single-user MIMO system aided by one IRS, and proposed an alternating optimization algorithm for finding a local optimum. \cite{huang2019ee} tackled the energy efficiency maximization problem using gradient descent approach and sequential fractional programming. Besides the passive beamforming design, there were also works focusing on IRS channel estimation (e.g., \cite{wang2020estimation,he2020estimation}) and deployment strategy (e.g., \cite{zhang2020mac,zhang2020region}).

However, the above IRS works only studied the scenario with one IRS or multiple non-cooperative IRSs (each independently serving its associated users). There is a lack of study for the inter-IRS reflection channel and the cooperative passive beamforming between IRSs. In \cite{han2020cooperative}, we made the first attempt to study a double-IRS aided wireless communication system with single-antenna BS/user. By assuming a rank-one inter-IRS reflection channel, an $M^4$-fold \emph{cooperative passive beamforming gain} can be achieved by the double-reflection link, with $M$ denoting the total number of IRS elements. This leads to a significantly increased achievable rate as compared to the single-IRS aided system with only an $M^2$-fold passive beamforming gain \cite{zhang2020tutorial}, especially when $M$ is large.

In this paper, we substantially extend this line of research by proposing a cooperative double-IRS aided MIMO system (as illustrated in Fig.~\ref{mimo} later), where one IRS is deployed near a multi-antenna BS and the other IRS is deployed near a multi-antenna user, for assisting the BS-user downlink communication. Note that the IRS has been shown extremely helpful for improving the MIMO channel capacity by boosting its rank via introducing reflection link. With one double-reflection link through two IRSs (BS-IRS$1$-IRS$2$-user) and two single-reflection links each via one IRS (BS-IRS$1$-user and BS-IRS$2$-user), the double-IRS aided MIMO system is further anticipated to achieve a more obvious rank (spatial multiplexing) gain comparing with the single-IRS aided MIMO system. This motivates our investigation in this paper to not only seek the cooperative passive beamforming gain but also the potential spatial multiplexing gain achievable by the double-IRS aided MIMO system. To this end, we consider a line-of-sight (LoS) propagation environment where the involved channels in Fig.~\ref{mimo} are generally of rank-one. Note that this condition can be practically achieved by properly deploying the two IRSs.

The main contributions of this paper are summarized as follows.
\begin{itemize}
\item{We are the first to study the capacity maximization problem of the double-IRS aided MIMO system, by jointly optimizing the transmit covariance matrix and the passive beamforming matrices of the two IRSs. This new optimization problem is non-convex and much more difficult than the single-IRS aided MIMO system (e.g., \cite{zhang2020capacity}). Nevertheless, we provide a tractable characterization of the LoS channels, and successfully derive a passive beamforming structure that can simultaneously maximize the power gains of the double-reflection link and two single-reflection links. Based on this, we transform and simplify the original problem, and propose a novel low-complexity algorithm by iteratively optimizing the transmit covariance matrix and the common phase shifts of the two IRSs in closed-form. Our algorithm's complexity is independent with the total number of IRS elements thanks to the exploitation of the LoS channel characteristics.}
\item{Next, we show that the rank of the double-IRS aided MIMO channel can be improved to two (as compared to the single-IRS aided MIMO channel which is always of rank-one), and analytically provide the explicit conditions that correspond to rank-two or rank-one MIMO channel. Moreover, by considering the case with two antennas at the BS or user, we derive the closed-form channel capacities for rank-two MIMO channel (with asymptotically large transmit power) and for rank-one MIMO channel. Based on the above, we analyze the capacity scaling orders with respect to an asymptotically large number of IRS elements or transmit power, which are shown to significantly outperform those of the conventional single-IRS aided MIMO system, thanks to the cooperative passive beamforming gain brought by the double-reflection link and the spatial multiplexing gain harvested from the two single-reflection links, respectively.}
\item{Finally, we present extensive numerical results. We show that by exploiting the LoS channel properties, our proposed algorithm can achieve a desirable performance with low computational time. We further validate our capacity scaling analysis, and reveal that the double-IRS aided MIMO system significantly outperforms its single-IRS counterpart when the total number of IRS elements or the transmit power is not small. In addition, we draw useful insights into the optimal IRS deployment for maximizing the channel capacity.}
\end{itemize}

The rest of this paper is organized as follows. In Section II, we introduce the system model and formulate the capacity maximization problem for the double-IRS aided MIMO system. In Section III, we provide a tractable characterization of the involved LoS channels. In Section VI, we propose a low-complexity algorithm to solve the considered problem. In Section V, we derive the closed-form channel capacity and conduct the capacity scaling analysis. In Section VI, we present numerical results for verifying the performance of our proposed algorithm and our considered double-IRS system. In Section VII, we conclude this paper.

\emph{Notations:} $|z|$, $z^*$, $\arg\{z\}$ denote the absolute value, conjugate, and angle of a complex number $z$. $\max(x,y)$ denotes the maximum between two real numbers $x$ and $y$. $\mathbb{C}$ denotes the space of complex numbers, $\mathbb{R}$ denotes the space of real numbers, while $\mathbb{Z}$ denotes the space of integers. Vectors and matrices are denoted by boldface lower-case letters and boldface upper-case letters, respectively. $(\cdot)^T$ denotes the transpose operation, while $(\cdot)^H$ denotes the conjugate transpose operation. For a vector $\mathbf{x}$, $\|\mathbf{x}\|$ and $(\mathbf{x})_{m}$ denote its $l_2$-norm and $k$th entry, respectively. For an arbitrary-size matrix $\mathbf{X}$, $\rank(\mathbf{X})$ and $(\mathbf{X})_{m,n}$ denote its rank and $(m,n)$th entry, respectively. $\diag\{x_1,\cdots,x_M\}$ denotes an $M\times M$ diagonal matrix with $x_1,\cdots,x_M$ being the diagonal elements. $\mathbf{I}_M$ denotes an $M\times M$ identity matrix, and $\mathbf{0}$ denotes an all-zero matrix with appropriate dimension. For a square matrix $\mathbf{S}$, $\det(\mathbf{S})$, $\tr(\mathbf{S})$, and $\mathbf{S}^{-1}$ denote its determinant, trace, and inverse, respectively, and $\mathbf{S}\succeq\mathbf{0}$ means that $\mathbf{S}$ is positive semi-definite. $j=\sqrt{-1}$ denotes the imaginary unit. $\mathbb{E}\{\cdot\}$ denotes the statistical expectation. $\mathcal{O}(\cdot)$ denotes the standard big-O notation.

\section{System Model and Problem Formulation}

\begin{figure}[t]
\centering
\includegraphics[width=4.0in]{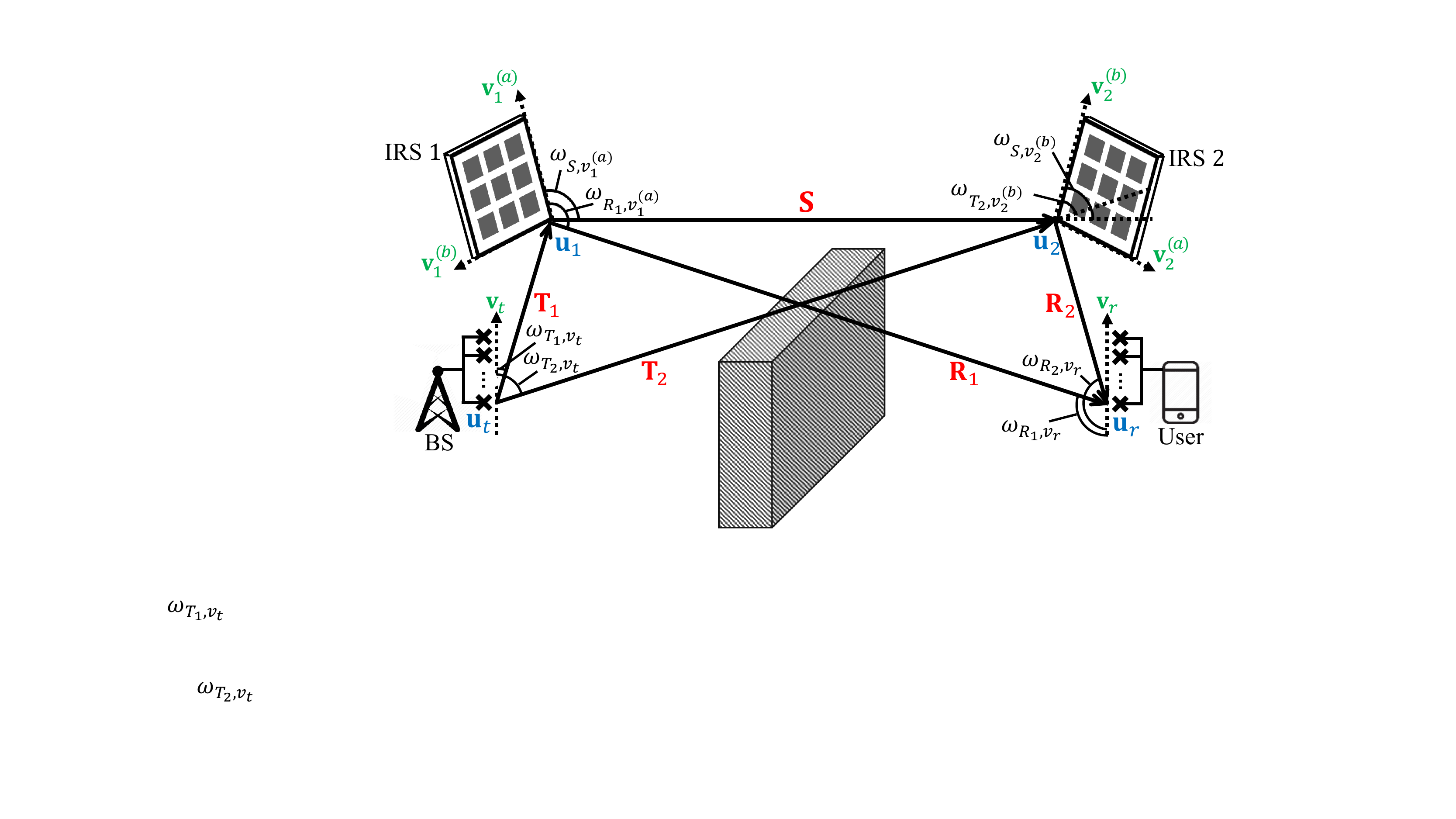}
\caption{A MIMO wireless communication system aided by two cooperative IRSs.}
\vspace{-1.5em}
\label{mimo}
\end{figure}

We study a MIMO downlink communication system with $N_t\geq1$ antennas at the BS and $N_r\geq1$ antennas at the user, with $\mathbf{u}_t\in\mathbb{R}^{3\times1}$ and $\mathbf{u}_r\in\mathbb{R}^{3\times1}$ denoting the locations of the BS and the user under a three-dimensional (3D) Cartesian coordinate system, respectively.\footnote{All the results can be easily extended to the multi-user setup, where different users take turns to be served in different time slots by applying our solution independently.} We consider the challenging scenario where the direct BS-user link is blocked by obstacles (e.g., high buildings for urban scenario and hills for suburban scenario), which is thus weak and negligible. We propose to deploy two cooperative IRSs to assist the communication in such a heavy-blockage scenario, as shown in Fig.~\ref{mimo}. For minimizing the path loss of the reflection links, we place IRS $1$ near the BS with its location denoted by $\mathbf{u}_1\in\mathbb{R}^{3\times1}$, and IRS $2$ near the user with its location denoted by $\mathbf{u}_2\in\mathbb{R}^{3\times1}$ as in \cite{han2020cooperative}. We are given a total number of $M\geq2$ passive reflecting elements as our budget, where IRS $i\in\{1,2\}$ is equipped with $M_i$ elements with $M_1+M_2=M$. For ease of exposition, we further define $\mathcal{N}_t=\{1,\cdots,N_t\}$ and $\mathcal{N}_r=\{1,\cdots,N_r\}$ as the sets containing all the antennas at the BS and the user, respectively, while $\mathcal{M}_i=\{1,\cdots,M_i\}$ as the set containing all the reflecting elements at IRS $i\in\{1,2\}$.

We denote $\mathbf{T}_i\in\mathbb{C}^{M_i\times N_t}$ as the channel matrix from the BS to IRS $i$, $\mathbf{R}_i\in\mathbb{C}^{N_r\times M_i}$ as the channel matrix from IRS $i$ to the user, for $i\in\{1,2\}$, and $\mathbf{S}\in\mathbb{C}^{M_2\times M_1}$ as the channel matrix from IRS $1$ to IRS $2$, respectively. We denote $\mathbf{\Phi}_{i}=\diag\{\phi_{i,1},\cdots,\phi_{i,M_i}\}\in\mathbb{C}^{M_i\times M_i}$ as the passive beamforming matrix of IRS $i\in\{1,2\}$, where $\phi_{i,m_i}$ denotes the reflection coefficient of element $m_i\in \mathcal{M}_i$ at IRS $i$. We allow each IRS to be equipped with a smart controller, which can intelligently adjust the reflection coefficients of its elements, i.e., $\phi_{i,m_i}$'s, to alter the effective BS-user MIMO channel for improving the communication performance. For achieving maximum reflection, we further set $|\phi_{i,m_i}|=1$ for $i\in\{1,2\}$ and $m_i\in\mathcal{M}_i$. Under the above setup, the effective BS-user MIMO channel aided by two cooperative IRSs is modelled as
\vspace{-0.6em}
\begin{equation}
\vspace{-0.6em}
\begin{aligned}
\mathbf{H} = \mathbf{R}_1\mathbf{\Phi}_1\mathbf{T}_1+\mathbf{R}_2\mathbf{\Phi}_2\mathbf{T}_2+\mathbf{R}_2\mathbf{\Phi}_2\mathbf{S}\mathbf{\Phi}_1\mathbf{T}_1,
\label{1}
\end{aligned}
\end{equation}which is the superposition of two \emph{single-reflection} links each via an individual IRS (BS-IRS$1$-user and BS-IRS$2$-user) and one \emph{double-reflection} link through both IRSs (BS-IRS$1$-IRS$2$-user).\footnote{Note that the double-reflection link reflected by IRS $2$ and then by IRS $1$ is very weak due to the high path loss, so are the links reflected by both IRSs more than twice. Thus, they can be neglected.} We further assume that the two IRSs are deployed such that all the involved channels ($\mathbf{T}_i$'s, $\mathbf{R}_i$'s, and $\mathbf{S}$) follow the free-space LoS propagation model \cite{goldsmith2005wireless} and can be expressed as
\begin{equation}
\begin{aligned}
\left\{
             \begin{array}{ll}
             (\mathbf{T}_i)_{m_i,n_t}=\frac{\sqrt{\alpha}}{d_{T_i,m_i,n_t}}e^{\frac{-j2\pi}{\lambda}d_{T_i,m_i,n_t}},\ i\in\{1,2\},\ m_i\in\mathcal{M}_i,\ n_t\in\mathcal{N}_t\\
(\mathbf{S})_{m_2,m_1}=\frac{\sqrt{\alpha}}{d_{S,m_2,m_1}}e^{\frac{-j2\pi}{\lambda}d_{S,m_2,m_1}},\ m_2\in\mathcal{M}_2,\ m_1\in\mathcal{M}_1\\
(\mathbf{R}_i)_{n_r,m_i}=\frac{\sqrt{\alpha}}{d_{R_i,n_r,m_i}}e^{\frac{-j2\pi}{\lambda}d_{R_i,n_r,m_i}},\ i\in\{1,2\},\ n_r\in\mathcal{N}_r,\ m_i\in\mathcal{M}_i,
             \end{array}
\right.
\label{2}
\end{aligned}
\end{equation}where $\alpha$ is the channel power gain at the reference distance of $1$ meter (m), $\lambda$ is the carrier wavelength; $d_{T_i,m_i,n_t}$, $d_{S,m_2,m_1}$, and $d_{R_i,n_r,m_i}$ denote the distance between antenna $n_t$ at the BS and element $m_{i}$ at IRS $i$, the distance between element $m_1$ at IRS $1$ and element $m_2$ at IRS $2$, and the distance between element $m_i$ at IRS $i$ and antenna $n_r$ at the user, respectively. We assume that the above channels are known a priori at the BS, user, and two IRSs, based on their geometric relationship.\footnote {In practice, such LoS channels can also be estimated by mounting each IRS with low-cost sensor arrays and using the IRS smart controller to send pilot signals.}

Let $\mathbf{x}\in\mathbb{C}^{N_t\times1}$ denote the transmitted signal vector. The transmit covariance matrix is thus defined as $\mathbf{Q}=\mathbb{E}\{\mathbf{x}\mathbf{x}^H\}$ with $\mathbf{Q}\succeq\mathbf{0}$ and $\tr(\mathbf{Q})\leq P$, where $P$ denotes the maximum transmit power of the BS. The received signal vector $\mathbf{y}\in\mathbb{C}^{N_r\times1}$ is given by
\vspace{-0.6em}
\begin{equation}
\vspace{-0.6em}
\begin{aligned}
\mathbf{y} = \mathbf{H}\mathbf{x}+\mathbf{z},
\label{3}
\end{aligned}
\end{equation}where $\mathbf{z}\sim\mathcal{CN}(0,\sigma^2\mathbf{I}_{N_r})$ denotes the independent circularly symmetric complex Gaussian (CSCG) noise vector at the user receiver, with $\sigma^2$ denoting the average noise power. According to \cite{goldsmith2005wireless}, the MIMO channel capacity is given by
\begin{equation}
\vspace{-0.2em}
\begin{aligned}
C = \underset{\mathbf{Q}:\tr(\mathbf{Q})\leq P, \mathbf{Q}\succeq \mathbf{0}}{\max}\log_2\det\bigg(\mathbf{I}_{N_r}+\frac{1}{\sigma^2}\mathbf{H}\mathbf{Q}\mathbf{H}^H\bigg).
\label{4}
\end{aligned}
\end{equation}

In this paper, we aim to maximize the channel capacity of our considered double-IRS aided MIMO system, by jointly optimizing the transmit covariance matrix $\mathbf{Q}$ and the passive beamforming matrices of the two IRSs $\{\mathbf{\Phi}_1,\mathbf{\Phi}_2\}$. The optimization problem is formulated as
\begin{equation}
\begin{aligned}
\!\!\!\!\!\!\!\!\!\!\!\!\!\!\!\!\!\!\!\!\!\text{(P1)}\ \ &\underset{\mathbf{\Phi}_1,\mathbf{\Phi}_2,\mathbf{Q}}{\max}\log_2\det\bigg(\mathbf{I}_{N_r}+\frac{1}{\sigma^2}\mathbf{H}\mathbf{Q}\mathbf{H}^H\bigg)
\label{5}
\end{aligned}
\end{equation}
\begin{equation}
\begin{aligned}
\ \ \ \ \ \ \ \text{s.t.}\ \ \ \ \mathbf{\Phi}_i=\diag\{\phi_{i,1},\cdots,\phi_{i,M_i}\},\ i\in\{1,2\}
\label{6}
\end{aligned}
\end{equation}
\begin{equation}
\begin{aligned}
\ \ \ \ \ \ \ \ \ |\phi_{i,m_i}|=1,\ i=\{1,2\},\ m_i\in\mathcal{M}_i
\label{7}
\end{aligned}
\end{equation}
\begin{equation}
\begin{aligned}
\!\!\!\!\!\!\!\!\!\!\!\!\!\!\!\!\!\!\!\!\!\!\!\!\!\!\!\!\!\!\!\!\tr(\mathbf{Q})\leq P
\label{8}
\end{aligned}
\end{equation}
\begin{equation}
\begin{aligned}
\!\!\!\!\!\!\!\!\!\!\!\!\!\!\!\!\!\!\!\!\!\!\!\!\!\!\!\!\!\!\!\!\mathbf{Q}\succeq\mathbf{0}.
\label{9}
\end{aligned}
\end{equation}Note that (P1) is a non-convex optimization problem given the non-concave objective function in \eqref{5} and the non-convex unit-modulus constraints in \eqref{7}. The complexity of exhaustive search for the optimal solution to (P1) grows exponentially with the total number of IRS elements $M$. Due to the coupling between the passive beamforming matrices of the two IRSs in (P1)'s objective function, (P1) is also more challenging than the MIMO channel capacity maximization problem studied in \cite{zhang2020capacity} with a single IRS. One may want to extend the alternating optimization algorithm in \cite{zhang2020capacity} to obtain a locally optimal solution to (P1), by iteratively optimizing the transmit covariance matrix and the reflection coefficients at the two IRSs. However, the involved complexity increases with $M$, and this approach does not exploit the unique structure of the LoS channels in this paper. Alternatively, we will present a more tractable characterization of the LoS channels in \eqref{2}, based on which we will develop a novel low-complexity algorithm for solving (P1).

For ease of reading, Table I summarizes the main symbol notations used in this paper and their physical meanings.

\begin{table*}[t]
\scriptsize
\newcommand{\tabincell}[2]{\begin{tabular}{@{}#1@{}}#2\end{tabular}}
\caption{Symbols and their physical meanings}
\centering
\begin{tabular}{|c|c||c|c|}
\hline
Symbols & Physical meanings & Symbols & Physical meanings\\
\hline
\hline
$\alpha$ & Channel power gain at reference distance of $1$ m & $P$ & Maximum transmit power of the BS\\
\hline
$\lambda$ & Carrier wavelength & $\sigma^2$ & Average noise power\\
\hline
$n_t/N_t$ & Antenna $n_t$/number of antennas at the BS & $K$ & Rank of $\mathbf{H}$\\
\hline
$n_r/N_r$ & Antenna $n_r$/number of antennas at the user & $\delta_k$ & $k$th singular value of $\mathbf{H}$\\
\hline
$m_i/M_i$ & \!\!Element $m_i$/number of elements at IRS $i$\!\! & $P_k$ & Transmit power allocated to $\delta_k$\\
\hline
$M$ & Total number of IRS elements & $d_{T_i}$ & Distance between the BS and IRS $i$\\
\hline
$l_n$ & Antenna spacing & $d_{S}$ & Distance between IRS $1$ and IRS $2$\\
\hline
$l_m$ & IRS element spacing & $d_{R_i}$ & Distance between IRS $i$ and the user\\
\hline
$\mathbf{u}_t$ & Location of the BS & $\mathbf{t}_{iL}/\mathbf{t}_{iR}$ & Array response of $\mathbf{T}_i$ at IRS $i$/the BS\\
\hline
$\mathbf{u}_r$ & Location of the user & $\mathbf{s}_{L}/\mathbf{s}_{R}$ & Array response of $\mathbf{S}$ at IRS $2$/IRS $1$\\
\hline
$\mathbf{u}_i$ & Location of IRS $i$ & $\mathbf{r}_{iL}/\mathbf{r}_{iR}$ & Array response of $\mathbf{R}_i$ at the user/IRS $i$\\
\hline
$\mathbf{v}_t$ & Base direction of the BS's antenna array & $\rho_{N_t}(\Theta_t)$ & Correlation between $\mathbf{t}_{1R}$ and $\mathbf{t}_{2R}$\\
\hline
$\mathbf{v}_r$ & Base direction of the user's antenna array & $\rho_{N_r}(\Theta_r)$ & Correlation between $\mathbf{r}_{1L}$ and $\mathbf{r}_{2L}$\\
\hline
$\mathbf{v}_i^{(a)}/\mathbf{v}_i^{(b)}$ & $1$st/$2$rd base direction of IRS $i$ & \tabincell{c}{$\omega_{T_i,v_t}/\omega_{T_i,v_i^{(a)}}/$\\$\omega_{T_i,v_i^{(b)}}$} & Angle between $\mathbf{T}_i$ and $\mathbf{v}_t$/$\mathbf{v}_i^{(a)}$/$\mathbf{v}_i^{(b)}$\\
\hline
$m_i^{(a)}/M_i^{(a)}$ & \tabincell{c}{Element $m_i^{(a)}$/number of elements in\\the $1$st base direction of IRS $i$} & $\omega_{S,v_i^{(a)}}/\omega_{S,v_i^{(b)}}$ & Angle between $\mathbf{S}$ and $\mathbf{v}_i^{(a)}$/$\mathbf{v}_i^{(b)}$\\
\hline
$m_i^{(b)}/M_i^{(b)}$ & \tabincell{c}{Element $m_i^{(b)}$/number of elements in\\the $2$rd base direction of IRS $i$} & \tabincell{c}{$\omega_{R_i,v_r}/\omega_{R_i,v_i^{(a)}}/$\\$\omega_{R_i,v_i^{(b)}}$} & Angle between $\mathbf{R}_i$ and $\mathbf{v}_r$/$\mathbf{v}_i^{(a)}$/$\mathbf{v}_i^{(b)}$\\
\hline
$\mathbf{T}_i$ & Channel matrix from the BS to IRS $i$ & $\beta_a$ & Reference phase of BS-IRS$1$-user link\\
\hline
$\mathbf{S}$ & Channel matrix from IRS $1$ to IRS $2$ & $\beta_b$ & Reference phase of BS-IRS$2$-user link\\
\hline
$\mathbf{R}_i$ & Channel matrix from IRS $i$ to the user & $\beta_c$ & \!Reference phase of BS-IRS$1$-IRS$2$-user link\!\\
\hline
$\mathbf{\Phi}_i$ & Passive beamforming matrix of IRS $i$ & $\gamma_i$ & Common phase shift of IRS $i$\\
\hline
$\phi_{i,m_i}$ & \!Reflection coefficient of element $m_i$ at IRS $i$\! & $|a(\mathbf{\Phi}_1)|$ & Power gain of BS-IRS$1$-user link\\
\hline
$\mathbf{H}$ & Effective channel from the BS to the user & $|b(\mathbf{\Phi}_2)|$ & Power gain of BS-IRS$2$-user link\\
\hline
$\mathbf{Q}$ & Transmit covariance matrix & $\!\!|c(\!\mathbf{\Phi}_1\!,\!\mathbf{\Phi}_2\!)|/|\tilde{c}(\!\mathbf{\Phi}_1\!,\!\mathbf{\Phi}_2\!)|\!\!$ & Power gain of BS-IRS$1$-IRS$2$-user link\\
\hline
\end{tabular}
\vspace{-0.5em}
\end{table*}

\newpage

\section{Tractable LoS Channel Characterization}

In this section, we provide a tractable characterization of the LoS channels in \eqref{2} and then that of the effective BS-user MIMO channel $\mathbf{H}$ in \eqref{1}.

Without loss of generality, we assume uniform linear array (ULA) for the BS/user and uniform rectangular array (URA) for the two IRSs, as shown in Fig.~\ref{mimo}.\footnote{The results of this paper are also applicable to any other antenna/reflecting element configurations by considering their corresponding array response vectors.} Specifically, we place the antennas at the BS and the user on lines along base directions $\mathbf{v}_t\in\mathbb{R}^{3\times1}$ and $\mathbf{v}_r\in\mathbb{R}^{3\times1}$, respectively, with $\|\mathbf{v}_t\|=\|\mathbf{v}_r\|=1$. We place the reflecting elements at IRS $i\in\{1,2\}$ on lines along two orthogonal base directions $\mathbf{v}_{i}^{(a)}\in \mathbb{R}^{3\times 1}$ and $\mathbf{v}_{i}^{(b)}\in \mathbb{R}^{3\times 1}$, respectively, with $\|\mathbf{v}_{i}^{(a)}\|=\|\mathbf{v}_{i}^{(b)}\|=1$. We denote the position of any particular element $m_i$ at IRS $i$ as $(m_{i}^{(a)},m_{i}^{(b)})$, which tells the indices in IRS $i$'s first and second base directions. Here, we have $m_{i}^{(a)}\in\{0,\cdots,M_{i}^{(a)}-1\}$ and $m_{i}^{(b)}\in\{0,\cdots,M_{i}^{(b)}-1\}$ with $M_i=M_{i}^{(a)}M_{i}^{(b)}$. Thus, we have the unique mapping between $m_i$ and $(m_{i}^{(a)},m_{i}^{(b)})$, i.e., $m_i=1+m_{i}^{(a)}+m_{i}^{(b)}M_{i}^{(a)}\in\{1,\cdots,M_{i}\}$.

We reasonably consider that the link distances of $\mathbf{T}_i$'s, $\mathbf{R}_i$'s, and $\mathbf{S}$ are much larger than the sizes of the BS's and user's antenna arrays as well as the sizes of the two IRSs, thus these LoS channels follow the far-field LoS channel model of rank-one as in \cite{gesbert2002outdoor}. Similar to \cite{han2020cooperative}, we rewrite $\mathbf{T}_i$ in \eqref{2} as
\begin{equation}
\vspace{-0.6em}
\begin{aligned}
\mathbf{T}_i=\frac{\sqrt{\alpha}}{d_{T_i}}e^{\frac{-j2\pi d_{T_i}}{\lambda}}\mathbf{t}_{iL}\mathbf{t}_{iR}^T,\ i\in\{1,2\},
\label{10}
\end{aligned}
\end{equation}which is the product of the path loss $\frac{\sqrt{\alpha}}{d_{T_i}}$, the reference phase $e^{\frac{-j2\pi d_{T_i}}{\lambda}}$, and the two array responses $\mathbf{t}_{iL}\in\mathbb{C}^{M_i\times1}$ and $\mathbf{t}_{iR}\in\mathbb{C}^{N_t\times1}$. Specifically, $d_{T_i}=\|\mathbf{u}_i-\mathbf{u}_t\|$ denotes the distance between (antenna $1$ at) the BS and (element $1$ at) IRS $i$,\footnote{This is because the link distance is much larger than the antenna/element spacing, and the distance between each pair of BS antenna and IRS element is almost the same and can be well approximated by $d_{T_i}$ when calculating the path loss.} and
\vspace{-0.4em}
\begin{equation}
\begin{aligned}
(\mathbf{t}_{iL})_{m_i}\!=\!\exp\!\bigg(\!\frac{-j2\pi}{\lambda}\Big(m_{i}^{(a)}l_{m}\cos\big(\omega_{T_i,v_i^{(a)}}\big)\!+\!m_{i}^{(b)}l_{m}\cos\big(\omega_{T_i,v_i^{(b)}}\big)\Big)\bigg),\ i\!\in\!\{1,2\},\ m_i\!\in\!\mathcal{M}_i,
\label{11}
\end{aligned}
\end{equation}
\vspace{-1.5em}
\begin{equation}
\vspace{-0.6em}
\begin{aligned}
(\mathbf{t}_{iR})_{n_t}=\exp\bigg(\frac{j2\pi}{\lambda}(n_{t}-1)l_{n}\cos\big(\omega_{T_i,v_t}\big)\bigg),\ i\in\{1,2\},\ n_t\in\mathcal{N}_t,
\label{12}
\end{aligned}
\end{equation}with $l_m$ denoting the IRS element spacing and $l_n$ denoting the antenna spacing. Note that $\omega_{T_i,v_t}$ denotes the angle between the direction of $\mathbf{T}_i$ and the base direction $\mathbf{v}_t$, i.e., $\omega_{T_i,v_t}=\arccos\big(\frac{(\mathbf{u}_i-\mathbf{u}_t)^T\mathbf{v}_t}{\|\mathbf{u}_i-\mathbf{u}_t\|\|\mathbf{v}_t\|}\big)\in[0,\pi]$, as shown in Fig.~\ref{mimo}. Similar definitions hold for the angle $\omega_{T_i,v_i^{(a)}}$ between $\mathbf{T}_i$ and $\mathbf{v}_i^{(a)}$, as well as the angle $\omega_{T_i,v_i^{(b)}}$ between $\mathbf{T}_i$ and $\mathbf{v}_i^{(b)}$.

Similar to $\mathbf{T}_i$ in \eqref{10}, $\mathbf{R}_i$ in \eqref{2} can be rewritten as
\vspace{-0.4em}
\begin{equation}
\vspace{-0.4em}
\begin{aligned}
\mathbf{R}_i=\frac{\sqrt{\alpha}}{d_{R_i}}e^{\frac{-j2\pi d_{R_i}}{\lambda}}\mathbf{r}_{iL}\mathbf{r}_{iR}^T,\ i\in\{1,2\},
\label{13}
\end{aligned}
\end{equation}where $d_{R_i}=\|\mathbf{u}_r-\mathbf{u}_i\|$ denotes the distance between IRS $i$ and the user, and the two array responses $\mathbf{r}_{iL}\in\mathbb{C}^{N_r\times1}$ and $\mathbf{r}_{iR}\in\mathbb{C}^{M_i\times1}$ are given by
\vspace{-0.4em}
\begin{equation}
\begin{aligned}
(\mathbf{r}_{iL})_{n_r}=\exp\bigg(\frac{-j2\pi}{\lambda}(n_{r}-1)l_{n}\cos\big(\omega_{R_i,v_r}\big)\bigg),\ i\in\{1,2\},\ n_r\in\mathcal{N}_r,
\label{14}
\end{aligned}
\end{equation}
\vspace{-1.5em}
\begin{equation}
\vspace{-0.4em}
\begin{aligned}
(\mathbf{r}_{iR})_{m_i}\!=\!\exp\!\bigg(\!\frac{j2\pi}{\lambda}\Big(m_{i}^{(a)}l_{m}\cos\big(\omega_{R_i,v_i^{(a)}}\big)\!+\!m_{i}^{(b)}l_{m}\cos\big(\omega_{R_i,v_i^{(b)}}\big)\Big)\bigg),\ i\!\in\!\{1,2\},\ m_i\!\in\!\mathcal{M}_i.
\label{15}
\end{aligned}
\end{equation}

Finally, $\mathbf{S}$ in \eqref{2} is rewritten as
\vspace{-0.4em}
\begin{equation}
\vspace{-0.4em}
\begin{aligned}
\mathbf{S}=\frac{\sqrt{\alpha}}{d_{S}}e^{\frac{-j2\pi d_{S}}{\lambda}}\mathbf{s}_L\mathbf{s}_R^T,
\label{16}
\end{aligned}
\end{equation}where $d_{S}=\|\mathbf{u}_2-\mathbf{u}_1\|$ denotes the distance between IRS $1$ and IRS $2$, and the two array responses $\mathbf{s}_L\in\mathbb{C}^{M_2\times1}$ and $\mathbf{s}_R\in\mathbb{C}^{M_1\times1}$ are given by
\begin{equation}
\begin{aligned}
(\mathbf{s}_L)_{m_2}=\exp\bigg(\frac{-j2\pi}{\lambda}\Big(m_{2}^{(a)}l_m\cos\big(\omega_{S,v_2^{(a)}}\big) + m_{2}^{(b)}l_m\cos\big(\omega_{S,v_2^{(b)}}\big)\Big)\bigg),\ m_2\in\mathcal{M}_2,
\label{17}
\end{aligned}
\end{equation}
\vspace{-1.5em}
\begin{equation}
\vspace{-0.4em}
\begin{aligned}
(\mathbf{s}_R)_{m_1}=\exp\bigg(\frac{j2\pi}{\lambda}\Big(m_{1}^{(a)}l_m\cos\big(\omega_{S,v_1^{(a)}}\big) + m_{1}^{(b)}l_m\cos\big(\omega_{S,v_1^{(b)}}\big)\Big)\bigg),\ m_1\in\mathcal{M}_1.
\label{18}
\end{aligned}
\end{equation}

By combining \eqref{10}, \eqref{13}, and \eqref{16}, we can successfully rewrite the double-IRS aided MIMO channel in \eqref{1} as
\vspace{-0.4em}
\begin{equation}
\vspace{-0.4em}
\begin{aligned}
\mathbf{H}=\frac{\alpha\beta_a}{d_{R_1}d_{T_1}}\mathbf{r}_{1L}\mathbf{r}_{1R}^T\mathbf{\Phi}_1\mathbf{t}_{1L}\mathbf{t}_{1R}^T+\frac{\alpha\beta_b}{d_{R_2}d_{T_2}}\mathbf{r}_{2L}\mathbf{r}_{2R}^T\mathbf{\Phi}_2\mathbf{t}_{2L}\mathbf{t}_{2R}^T+\frac{\alpha^{3/2}\beta_c}{d_{R_2}d_{S}d_{T_1}}\mathbf{r}_{2L}\mathbf{r}_{2R}^T\mathbf{\Phi}_2\mathbf{s}_L\mathbf{s}_R^T\mathbf{\Phi}_1\mathbf{t}_{1L}\mathbf{t}_{1R}^T\nonumber
\end{aligned}
\end{equation}
\begin{equation}
\vspace{-0.4em}
\begin{aligned}
\!\!\!\!\!\!\!\!\!\!\!\!\!\!\!\!\!\!\!\!\!\!\!\!\!\!\!\!\!\!\!\!\!\!\!\!\!\!\!\!\!\!\!\!\!\!\!\!\!\!\!\!\!\!\!\!\!\!\!\!\!\!\!\!\!\!\!\!\!\!\!\!\!\!\!\!\!\!\!\!\!\!\!\!\!\!\!=a(\mathbf{\Phi}_1)\mathbf{r}_{1L}\mathbf{t}_{1R}^T+b(\mathbf{\Phi}_2)\mathbf{r}_{2L}\mathbf{t}_{2R}^T+c(\mathbf{\Phi}_1,\mathbf{\Phi}_2)\mathbf{r}_{2L}\mathbf{t}_{1R}^T,
\label{19}
\end{aligned}
\end{equation}where $\beta_a=e^{\frac{-j2\pi (d_{R_1}\!+d_{T_1})}{\lambda}}$, $\beta_b=e^{\frac{-j2\pi (d_{R_2}\!+d_{T_2})}{\lambda}}$, and $\beta_c=e^{\frac{-j2\pi (d_{R_2}\!+d_{S}+d_{T_1})}{\lambda}}$ denote the reference phases of the BS-IRS$1$-user link, BS-IRS$2$-user link, and BS-IRS$1$-IRS$2$-user link, repectively. For ease of characterizing $\mathbf{H}$, we further define
\vspace{-0.4em}
\begin{equation}
\vspace{-0.4em}
\begin{aligned}
\left\{
             \begin{array}{ll}
             a(\mathbf{\Phi}_1)\!=\!\frac{\alpha\beta_a}{d_{R_1}\!d_{T_1}}\mathbf{r}_{1R}^T\mathbf{\Phi}_1\mathbf{t}_{1L}\\
b(\mathbf{\Phi}_2)\!=\!\frac{\alpha\beta_b}{d_{R_2}\!d_{T_2}}\mathbf{r}_{2R}^T\mathbf{\Phi}_2\mathbf{t}_{2L}\\
c(\mathbf{\Phi}_1,\mathbf{\Phi}_2)\!=\!\frac{\alpha^{3/2}\beta_c}{d_{R_2}\!d_{S}d_{T_1}}(\mathbf{r}_{2R}^T\mathbf{\Phi}_2\mathbf{s}_L)(\mathbf{s}_R^T\mathbf{\Phi}_1\mathbf{t}_{1L}).
             \end{array}
\right.
\label{20}
\end{aligned}
\end{equation}
It is worth noting that $|a(\mathbf{\Phi}_1)|$, $|b(\mathbf{\Phi}_2)|$, and $|c(\mathbf{\Phi}_1,\mathbf{\Phi}_2)|$ determine the power gains of the BS-IRS$1$-user link, BS-IRS$2$-user link, and BS-IRS$1$-IRS$2$-user link, repectively.

\section{Proposed Low-complexity Solution to (P1)}

Based on the new tractable LoS channel characteristics in the last section, in this section, we aim to propose a low-complexity algorithm for solving (P1), to fit the practical scenario with a large total number of IRS elements $M$. Specifically, we first design a passive beamforming structure that can simultaneously maximize the power gains of the double-reflection link and two single-reflection links. Then, we transform the original problem (P1) to a simplified problem (P2) with only three optimization variables, i.e., the transmit covariance matrix and the common phase shifts of the two IRSs. We further derive the closed-form solutions to the three subproblems of (P2), for optimizing the transmit covariance matrix or one common phase shift with the other two variables being fixed. Finally, we propose an alternating optimization algorithm to obtain a locally optimal solution to (P2), by iteratively solving the above three subproblems.

Since the BS-IRS$1$ distance and the IRS$2$-user distance are much smaller than the IRS$1$-IRS$2$ distance and the BS-user distance, $\mathbf{S}$ is approximately parallel to $\mathbf{R}_1$ and $\mathbf{T}_2$, namely, $\omega_{R_1,v_1^{(a)}}\approx\omega_{S,v_1^{(a)}}$, $\omega_{R_1,v_1^{(b)}}\approx\omega_{S,v_1^{(b)}}$, $\omega_{T_2,v_2^{(a)}}\approx\omega_{S,v_2^{(a)}}$, and $\omega_{T_2,v_2^{(b)}}\approx\omega_{S,v_2^{(b)}}$. Hence, $\mathbf{t}_{2L}$ in \eqref{11} and $\mathbf{s}_L$ in \eqref{17} are almost the same, so are $\mathbf{r}_{1R}$ in \eqref{15} and $\mathbf{s}_R$ in \eqref{18}. Therefore, we have $c(\mathbf{\Phi}_1,\mathbf{\Phi}_2)\approx\check{c}(\mathbf{\Phi}_1,\mathbf{\Phi}_2)=\frac{\alpha^{3/2}\beta_c}{d_{R_2}\!d_{S}d_{T_1}}(\mathbf{r}_{2R}^T\mathbf{\Phi}_2\mathbf{t}_{2L})(\mathbf{r}_{1R}^T\mathbf{\Phi}_1\mathbf{t}_{1L})$, and \eqref{19} can be well approximated by
\vspace{-0.5em}
\begin{equation}
\vspace{-0.5em}
\begin{aligned}
\mathbf{H}\approx a(\mathbf{\Phi}_1)\mathbf{r}_{1L}\mathbf{t}_{1R}^T+b(\mathbf{\Phi}_2)\mathbf{r}_{2L}\mathbf{t}_{2R}^T+\check{c}(\mathbf{\Phi}_1,\mathbf{\Phi}_2)\mathbf{r}_{2L}\mathbf{t}_{1R}^T.
\label{21}
\end{aligned}
\end{equation}Consequently, $|a(\mathbf{\Phi}_1)|$, $|b(\mathbf{\Phi}_2)|$, and $|\check{c}(\mathbf{\Phi}_1,\mathbf{\Phi}_2)|$ can be simultaneously maximized by deploying the following passive beamforming structure:
\vspace{-0.5em}
\begin{equation}
\vspace{-0.5em}
\begin{aligned}
\left\{
             \begin{array}{ll}
             \phi_{1,m_1}=\gamma_1(\mathbf{t}_{1L})_{m_1}^*(\mathbf{r}_{1R})_{m_1}^*, \ \ m_1\in\mathcal{M}_1\\
\phi_{2,m_2}=\gamma_2(\mathbf{t}_{2L})_{m_2}^*(\mathbf{r}_{2R})_{m_2}^*, \ \ m_2\in\mathcal{M}_2,
             \end{array}
\right.
\label{22}
\end{aligned}
\end{equation}
where $\gamma_1$ with $|\gamma_1|=1$ and $\gamma_2$ with $|\gamma_2|=1$ denote the common phase shifts of IRS $1$ and IRS $2$, respectively. The maximum power gains of the single-reflection links via IRS $1$ and IRS $2$ as well as the double-reflection link are thus given by $|a(\mathbf{\Phi}_1)|\!=\!\frac{\alpha M_1}{d_{R_1}\!d_{T_1}}$, $|b(\mathbf{\Phi}_2)|\!=\!\frac{\alpha M_2}{d_{R_2}\!d_{T_2}}$, and $|\check{c}(\mathbf{\Phi}_1,\mathbf{\Phi}_2)|\!=\!\frac{\alpha^{3/2} M_1M_2}{d_{R_2}\!d_{S}d_{T_1}}$, respectively.

Inspired by the above, we propose to adopt the passive beamforming structure in \eqref{22}, and further optimize the common phase shifts $\gamma_1$ and $\gamma_2$ to maximize the MIMO channel capacity. In this case, the effective MIMO channel in \eqref{21} can be rewritten as
\vspace{-0.4em}
\begin{equation}
\vspace{-0.4em}
\begin{aligned}
\mathbf{H}=\frac{\alpha M_1\beta_a\gamma_1}{d_{R_1}\!d_{T_1}}\mathbf{r}_{1L}\mathbf{t}_{1R}^T+\frac{\alpha M_2\beta_b\gamma_2}{d_{R_2}\!d_{T_2}}\mathbf{r}_{2L}\mathbf{t}_{2R}^T+\frac{\alpha^{3/2} M_1M_2\beta_c\gamma_1\gamma_2}{d_{R_2}\!d_{S}d_{T_1}}\mathbf{r}_{2L}\mathbf{t}_{1R}^T,
\label{23}
\end{aligned}
\end{equation}and the optimization problem (P1) is reformulated as
\vspace{-0.5em}
\begin{equation}
\begin{aligned}
\!\!\!\!\!\!\!\!\!\!\!\!\!\!\!\!\!\!\!\!\!\text{(P2)}\ \ &\underset{\gamma_1,\gamma_2,\mathbf{Q}}{\max}\log_2\det\bigg(\mathbf{I}_{N_r}+\frac{1}{\sigma^2}\mathbf{H}\mathbf{Q}\mathbf{H}^H\bigg)
\label{24}
\end{aligned}
\end{equation}
\vspace{-2.0em}
\begin{equation}
\begin{aligned}
\!\!\!\!\!\!\!\!\!\!\!\!\!\!\!\!\!\!\!\!\!\!\!\!\!\!\!\!\!\!\!\!\!\text{s.t.}\ \ \ |\gamma_i|=1,\ i=\{1,2\}
\label{25}
\end{aligned}
\end{equation}
\vspace{-2.5em}
\begin{equation}
\begin{aligned}
\!\!\!\!\!\!\!\!\!\!\!\!\!\!\!\!\!\!\!\!\!\!\!\!\!\!\!\!\!\!\!\!\!\tr(\mathbf{Q})\leq P
\label{26}
\end{aligned}
\end{equation}
\vspace{-2.5em}
\begin{equation}
\vspace{-0.6em}
\begin{aligned}
\!\!\!\!\!\!\!\!\!\!\!\!\!\!\!\!\!\!\!\!\!\!\!\!\!\!\!\!\!\!\!\!\!\mathbf{Q}\succeq\mathbf{0}.
\label{27}
\end{aligned}
\end{equation}Note that our proposed approach transforms the original problem (P1) with $M+1$ optimization variables $\{\mathbf{Q}\}\bigcup\{\phi_{1,m_1}\}_{m_1=1}^{M_1}\bigcup\{\phi_{2,m_2}\}_{m_2=1}^{M_2}$ to a simplified problem (P2) with only $3$ optimization variables $\{\mathbf{Q},\gamma_1,\gamma_2\}$. Although (P2) is still a non-convex optimization problem, in the following, we first optimally solve the three subproblems of (P2) in closed-form, which aim to optimize one variable in $\{\mathbf{Q},\gamma_1,\gamma_2\}$ with the other two variables being fixed. We then present an efficient alternating optimization algorithm for obtaining a locally optimal solution to (P2), by iteratively solving the above three subproblems.

\subsection{Optimization of $\mathbf{Q}$ with Given $\gamma_1$ and $\gamma_2$}

First, we consider the subproblem of optimizing the transmit covariance matrix $\mathbf{Q}$ with given $\gamma_1$ and $\gamma_2$. Note that (P2) is a convex optimization problem about $\mathbf{Q}$. The truncated singular value decomposition (SVD) of $\mathbf{H}$ in \eqref{23} is given by $\mathbf{H}=\mathbf{U}\mathbf{\Delta}\mathbf{V}^H$, where $\mathbf{V}\in\mathbb{C}^{N_t\times K}$ with $K=\rank(\mathbf{H})$, and the optimal $\mathbf{Q}^\star$ is given in the lemma below by following the water-filling power allocation \cite{goldsmith2005wireless}.

\underline{\emph{Lemma 1}}: The optimal $\mathbf{Q}^\star$ with given $\gamma_1$ and $\gamma_2$ is
\vspace{-0.6em}
\begin{equation}
\vspace{-0.6em}
\begin{aligned}
\mathbf{Q}^\star=\mathbf{V}\diag\{P_1^\star,\cdots,P_K^\star\}\mathbf{V}^H,
\label{28}
\end{aligned}
\end{equation}where $P_k^\star$ is the optimal amount of transmit power allocated to the $k$th singular value, i.e., $P_k^\star=\max(\mu-\frac{\sigma^2}{\delta_k^2},0)$, $k=1,\cdots,K$, with $\delta_k=(\mathbf{\Delta})_{k,k}$ and $\mu$ satisfying $\sum_{k=1}^KP_k^\star=P$.

\subsection{Optimization of $\gamma_1$ with Given $\mathbf{Q}$ and $\gamma_2$}

Next, we aim to optimize the common phase shift of IRS $1$ denoted by $\gamma_1$ with given $\mathbf{Q}$ and $\gamma_2$. Note that the eigenvalue decomposition (EVD) of $\mathbf{Q}$ is given by $\mathbf{Q}=\mathbf{U}_Q\mathbf{\Sigma}_Q\mathbf{U}_Q^H$, where $\mathbf{U}_Q\in\mathbb{C}^{N_t\times N_t}$ and $\mathbf{\Sigma}_Q\in\mathbb{C}^{N_t\times N_t}$. For ease of exposition, we define $\mathbf{A}=\frac{\alpha M_1\beta_a}{d_{R_1}\!d_{T_1}}\mathbf{r}_{1L}\mathbf{t}_{1R}^T\mathbf{U}_Q\mathbf{\Sigma}_Q^{\frac{1}{2}}$, $\mathbf{B}=\frac{\alpha M_2\beta_b}{d_{R_2}\!d_{T_2}}\mathbf{r}_{2L}\mathbf{t}_{2R}^T\mathbf{U}_Q\mathbf{\Sigma}_Q^{\frac{1}{2}}$, $\mathbf{C}=\frac{\alpha^{3/2} M_1M_2\beta_c}{d_{R_2}\!d_{S}d_{T_1}}\mathbf{r}_{2L}\mathbf{t}_{1R}^T\mathbf{U}_Q\mathbf{\Sigma}_Q^{\frac{1}{2}}$, and the objective function of (P2) with respect to $\gamma_1$ can be rewritten as
\vspace{-0.5em}
\begin{equation}
\vspace{-0.5em}
\begin{aligned}
f_{\gamma_1}&=\log_2\det\bigg(\mathbf{I}_{N_r}+\frac{1}{\sigma^2}\mathbf{H}\mathbf{Q}\mathbf{H}^H\bigg)\\
&=\log_2\det\bigg(\mathbf{I}_{N_r}+\frac{1}{\sigma^2}\big(\gamma_1\mathbf{A}+\gamma_2\mathbf{B}+\gamma_1\gamma_2\mathbf{C}\big)\big(\gamma_1\mathbf{A}+\gamma_2\mathbf{B}+\gamma_1\gamma_2\mathbf{C}\big)^H\bigg)\\
&=\log_2\det\big(\mathbf{X}_{1}+\gamma_1\mathbf{Y}_{1}+\gamma_1^*\mathbf{Y}_{1}^H\big),
\label{29}
\end{aligned}
\end{equation}
where
\vspace{-0.3em}
\begin{equation}
\vspace{-0.3em}
\begin{aligned}
\mathbf{X}_{1}=\mathbf{I}_{N_r}+\frac{1}{\sigma^2}\Big(\mathbf{A}\mathbf{A}^H+\mathbf{B}\mathbf{B}^H+\mathbf{C}\mathbf{C}^H+\gamma_2\mathbf{C}\mathbf{A}^H+\gamma_2^*\mathbf{A}\mathbf{C}^H\Big),
\label{30}
\end{aligned}
\end{equation}and
\vspace{-0.3em}
\begin{equation}
\vspace{-0.3em}
\begin{aligned}
\mathbf{Y}_{1}=\frac{1}{\sigma^2}\Big(\gamma_2^*\mathbf{A}\mathbf{B}^H+\mathbf{C}\mathbf{B}^H\Big).
\label{31}
\end{aligned}
\end{equation}

Therefore, the subproblem with respect to $\gamma_1$ can be formulated as
\begin{equation}
\vspace{-0.5em}
\begin{aligned}
\vspace{-0.5em}
\text{(P2-$\gamma_1$)}\ \ \ &\underset{\gamma_1}{\max}\ \log_2\det\big(\mathbf{X}_{1}+\gamma_{1}\mathbf{Y}_{1}+\gamma_{1}^*\mathbf{Y}_{1}^H\big)\\
&\ \ \text{s.t.}\ |\gamma_{1}|=1.
\label{32}
\end{aligned}
\end{equation}Note that (P2-$\gamma_1$) has a similar structure as (P1-$m$) in \cite{zhang2020capacity}. Therefore, the optimal solution to (P2-$\gamma_1$) can be similarly derived as the optimal solution to (P1-$m$) in \cite{zhang2020capacity}, as shown in the following lemma.

\underline{\emph{Lemma 2}}: The optimal $\gamma_1^\star$ with given $\mathbf{Q}$ and $\gamma_2$ is
\vspace{-0.5em}
\begin{equation}
\vspace{-0.5em}
\begin{aligned}
\gamma_1^\star=\left\{
             \begin{array}{ll}
             e^{-j\arg\{\nu_{1}\}}, & \text{if}\ \ \tr\big(\mathbf{X}_{1}^{-1}\mathbf{Y}_{1}\big)\neq0\\
1, & \text{otherwise},
             \end{array}
\right.
\label{33}
\end{aligned}
\end{equation}where $\nu_{1}$ is the sole non-zero eigenvalue of $\mathbf{X}_{1}^{-1}\mathbf{Y}_{1}$.

\subsection{Optimization of $\gamma_{2}$ with Given $\mathbf{Q}$ and $\gamma_1$}

Finally, we focus on the subproblem to optimize the common phase shift of IRS $2$ denoted by $\gamma_2$ with given $\mathbf{Q}$ and $\gamma_1$. Similar to the previous subsection for optimizing $\gamma_1$, the subproblem for optimizing $\gamma_{2}$ is formulated as
\vspace{-0.2em}
\begin{equation}
\vspace{-0.5em}
\begin{aligned}
\text{(P2-$\gamma_2$)}\ \ \ &\underset{\gamma_{2}}{\max}\ \log_2\det\big(\mathbf{X}_{2}+\gamma_{2}\mathbf{Y}_{2}+\gamma_{2}^*\mathbf{Y}_{2}^H\big)\\
&\ \ \text{s.t.}\ \ |\gamma_{2}|=1,
\label{34}
\end{aligned}
\end{equation}where
\vspace{-0.3em}
\begin{equation}
\vspace{-0.3em}
\begin{aligned}
\mathbf{X}_{2}=\mathbf{I}_{N_r}+\frac{1}{\sigma^2}\Big(\mathbf{A}\mathbf{A}^H+\mathbf{B}\mathbf{B}^H+\mathbf{C}\mathbf{C}^H+\gamma_1\mathbf{C}\mathbf{B}^H+\gamma_1^*\mathbf{B}\mathbf{C}^H\Big),
\label{35}
\end{aligned}
\end{equation}and
\vspace{-0.3em}
\begin{equation}
\vspace{-0.3em}
\begin{aligned}
\mathbf{Y}_{2}=\frac{1}{\sigma^2}\Big(\gamma_1^*\mathbf{B}\mathbf{A}^H+\mathbf{C}\mathbf{A}^H\Big).
\label{36}
\end{aligned}
\end{equation}Therefore, the optimal solution to (P2-$\gamma_2$) is given by
\vspace{-0.5em}
\begin{equation}
\vspace{-0.5em}
\begin{aligned}
\gamma_{2}^\star=\left\{
             \begin{array}{ll}
             e^{-j\arg\{\nu_{2}\}}, & \text{if}\ \ \tr\big(\mathbf{X}_{2}^{-1}\mathbf{Y}_{2}\big)\neq0\\
1, & \text{otherwise},
             \end{array}
\right.
\label{37}
\end{aligned}
\end{equation}where $\nu_{2}$ is the sole non-zero eigenvalue of $\mathbf{X}_{2}^{-1}\mathbf{Y}_{2}$.

\subsection{Overall Algorithm}

With all the three subproblems optimally solved in closed-form, we are ready to present the overall algorithm to solve (P2). Specifically, we initialize with $\gamma_1=\gamma_2=1$. Then, we iteratively optimize the transmit covariance matrix $\mathbf{Q}$ or one common phase shift $\gamma_{i}$ at each time, with the other two variables being fixed. The convergence is reached if the relative increment of (P2)'s objective function does not exceed a threshold $\epsilon>0$ by optimizing any variable in $\{\mathbf{Q},\gamma_1,\gamma_2\}$. The overall algorithm is summarized in Algorithm 1. Note that each subproblem is optimally solved in closed-form, and the optimization variables are not coupled in (P2)'s constraints. Therefore, Algorithm 1 is guaranteed to converge to at least a locally optimal solution to (P2), since the algorithm yields a non-decreasing objective function over iterations, which is upper-bounded by a finite value. It is worth noting that based on the obtained solution to (P2), a suboptimal solution to (P1) is automatically obtained by substituting $\gamma_1^\star$ and $\gamma_2^\star$ into \eqref{22}.

As compared to the very recent algorithm for capacity maximization in \cite{zhang2020capacity} whose complexity increases fast with the total number of IRS elements $M$, the complexity of our proposed algorithm is \emph{independent with $M$}, which greatly saves the computational time especially for the practical scenario with large $M$.

\begin{algorithm}[t]
\DontPrintSemicolon
  
  \KwInput{$\mathbf{T}_1$, $\mathbf{T}_2$, $\mathbf{R}_1$, $\mathbf{R}_2$, $\mathbf{S}$, $P$, $\sigma^2$}
  \KwOutput{$\mathbf{Q}^\star$, $\gamma_1^\star$, $\gamma_2^\star$}
  Initialize $\gamma_{1}=1$ and $\gamma_{2}=1$.\\
  {\bf{Repeat}}\\
  \ \ \ Obtain the optimal $\mathbf{Q}^\star$ in \eqref{28}.\\
  \ \ \ Obtain the optimal $\gamma_1^\star$ in \eqref{33}.\\
  \ \ \ Obtain the optimal $\gamma_2^\star$ in \eqref{37}.\\
  {\bf{Until}} The relative increment of the objective function in \eqref{24} does not exceed a threshold $\epsilon>0$ by optimizing any variable in $\{\mathbf{Q},\gamma_1,\gamma_2\}$.
   		
\caption{Alternating optimization algorithm to (P2)}
\end{algorithm}

\section{Capacity Scaling Analysis for Optimized Double-IRS Aided MIMO System}

In this section, we derive the capacity scaling orders of the optimized double-IRS aided MIMO system with respect to asymptotically large total number of IRS elements $M$ or transmit power $P$, for drawing more insights about its asymptotic capacity performance. Specifically, we first analytically derive the explicit conditions that the effective MIMO channel $\mathbf{H}$ in \eqref{21} is of rank-two or rank-one. Then, we analyze the MIMO channel capacities as well as the scaling orders for rank-two $\mathbf{H}$ and rank-one $\mathbf{H}$, respectively.

\subsection{Rank Analysis of Double-IRS Aided MIMO Channel}

First, we characterize the rank of the double-IRS aided MIMO channel $\mathbf{H}$. As we can see from \eqref{21}, the double-reflection link can be combined with one of the single-reflection links as one rank-one matrix, thus $K=\rank(\mathbf{H})$ is upper-bounded as
\vspace{-0.5em}
\begin{equation}
\vspace{-0.5em}
\begin{aligned}
K&\leq\rank\big(a(\mathbf{\Phi}_1)\mathbf{r}_{1L}\mathbf{t}_{1R}^T\big)+\rank\Big(\mathbf{r}_{2L}\big(b(\mathbf{\Phi}_2)\mathbf{t}_{2R}^T+\check{c}(\mathbf{\Phi}_1,\mathbf{\Phi}_2)\mathbf{t}_{1R}^T\big)\Big)=2.
\label{38}
\end{aligned}
\end{equation}Based on the LoS channel characterization, we express the correlation between the two array responses at the BS in the BS-IRS$1$ and BS-IRS$2$ channels, $\mathbf{t}_{1R}$ and $\mathbf{t}_{2R}$ in \eqref{12}, as
\vspace{-0.5em}
\begin{equation}
\vspace{-0.5em}
\begin{aligned}
\rho_{N_t}(\Theta_t)=\frac{1}{N_t}\big|\mathbf{t}_{2R}^H\mathbf{t}_{1R}\big|=\frac{1}{N_t}\bigg|\sum_{n_t=1}^{N_t}e^{-j2\pi(n_t-1)\Theta_{t}}\bigg|=\frac{\sin(\pi N_t\Theta_{t})}{N_t\sin(\pi \Theta_{t})}\in[0,1],
\label{39}
\end{aligned}
\end{equation}
where $\Theta_t=\frac{l_{n}}{\lambda}\big(\cos(\omega_{T_2,v_t})-\cos(\omega_{T_1,v_t})\big)$. Similarly, we express the correlation between the two array responses at the user in the IRS$1$-user and IRS$2$-user channels, $\mathbf{r}_{1L}$ and $\mathbf{r}_{2L}$ in \eqref{14}, as
\vspace{-0.5em}
\begin{equation}
\vspace{-0.5em}
\begin{aligned}
\rho_{N_r}(\Theta_r)=\frac{1}{N_r}\big|\mathbf{r}_{2L}^H\mathbf{r}_{1L}\big|=\frac{1}{N_r}\bigg|\sum_{n_r=1}^{N_r}e^{-j2\pi(n_r-1)\Theta_{r}}\bigg|=\frac{\sin(\pi N_r\Theta_{r})}{N_r\sin(\pi\Theta_{r})}\in[0,1],
\label{40}
\end{aligned}
\end{equation}
where $\Theta_{r}=\frac{l_{n}}{\lambda}\big(\cos(\omega_{R_1,v_r})-\cos(\omega_{R_2,v_r})\big)$. It can be observed from \eqref{39} and \eqref{40}, when $\rho_{N_t}(\Theta_t)=1$ or $\rho_{N_r}(\Theta_r)=1$, $\mathbf{H}$ is of rank-one. This is because $\rho_{N_t}(\Theta_t)=1$ means that $\mathbf{t}_{1R}$ and $\mathbf{t}_{2R}$ are the same, and the three rank-one matrices in \eqref{21} can be combined as one rank-one matrix. Similar analysis holds for $\rho_{N_r}(\Theta_r)=1$. While for the case of $\rho_{N_t}(\Theta_t)\neq1$ and $\rho_{N_r}(\Theta_r)\neq1$, $\mathbf{t}_{1R}$ and $\mathbf{t}_{2R}$ are linearly independent, so are $\mathbf{r}_{1L}$ and $\mathbf{r}_{2L}$, thus the equality in \eqref{38} holds and $\mathbf{H}$ is of rank-two. Therefore, we have the following proposition about the rank of $\mathbf{H}$.

\underline{\emph{Proposition 1}}: \ Under the LoS channels, the rank of the double-IRS aided MIMO channel $\mathbf{H}$ is given by
\begin{equation}
\vspace{-0.5em}
\begin{aligned}
K=\left\{
             \begin{array}{ll}
             2, & \text{if}\ \rho_{N_t}(\Theta_t)\neq1\ \text{and}\ \rho_{N_r}(\Theta_r)\neq1\\
1, & \text{otherwise}.
             \end{array}
\right.
\label{41}
\end{aligned}
\end{equation}It is worth noting that $\rho_{N_t}(\Theta_t)$ is periodic with period $1$ and takes the maximal value $1$ at $\Theta_t=0$, thus $\rho_{N_t}(\Theta_t)\neq1$ corresponds to $\Theta_t=\frac{l_{n}}{\lambda}\big(\cos(\omega_{T_2,v_t})-\cos(\omega_{T_1,v_t})\big)\notin\mathbb{Z}$. Also, the difference between $\cos(\omega_{T_1,v_t})$ and $\cos(\omega_{T_2,v_t})$ cannot be more than $2$. Similar analysis holds for $\rho_{N_r}(\Theta_r)$. Therefore, we have the following corollary.

\underline{\emph{Corollary 1}}: \ For the case with antenna spacing $l_n\leq\frac{\lambda}{2}$, the explicit conditions in \eqref{41} can be further simplified as
\vspace{-0.5em}
\begin{equation}
\vspace{-0.5em}
\begin{aligned}
K=\left\{
             \begin{array}{ll}
             2, & \text{if}\ \omega_{T_1,v_t}\neq\omega_{T_2,v_t}\ \text{and}\ \omega_{R_1,v_r}\neq\omega_{R_2,v_r}\\
1, & \text{otherwise},
             \end{array}
\right.
\label{42}
\end{aligned}
\end{equation}which means that $\mathbf{H}$ is of rank-two if the two angles-of-departure, $\omega_{T_1,v_t}$ and $\omega_{T_2,v_t}$, are different (as shown in Fig.~\ref{mimo}), so are the two angles-of-arrival, $\omega_{R_1,v_r}$ and $\omega_{R_2,v_r}$.

In the following, we analytically derive the MIMO channel capacities and the scaling orders of the double-IRS aided MIMO system for rank-two $\mathbf{H}$ and rank-one $\mathbf{H}$, respectively. For the purpose of exposition, we consider the setup with $N_t\geq 2$ antennas at the BS and $N_r=2$ antennas at the user, without compromising the spatial multiplexing gain, while all the results are also applicable to the setup with $N_t=2$ antennas at the BS and $N_r\geq2$ antennas at the user by leveraging the uplink-downlink duality. Later in Section VI, we will provide numerical results on the capacity scaling with arbitrary number of antennas at the BS/user.

\subsection{Capacity Scaling with Rank-two $\mathbf{H}$}

We first consider the case that $\mathbf{H}$ in \eqref{21} is of rank-two, i.e., when $\rho_{N_t}(\Theta_t)\neq1$ and $\rho_{N_r}(\Theta_r)\neq1$ as in \eqref{41}. The MIMO channel capacity can be expressed as
\vspace{-0.5em}
\begin{equation}
\vspace{-0.5em}
\begin{aligned}
C&=\sum_{k = 1}^{2}\log_2\bigg(1+\frac{P_k^\star\delta_k^2}{\sigma^2}\bigg),
\label{43}
\end{aligned}
\end{equation}where $P_k^\star$ is the optimal amount of transmit power allocated to the $k$th singular value of $\mathbf{H}$, $\delta_k>0$, $k=1,2$, as given in \eqref{28} with $K=2$. In this case, the capacity scaling order with respect to an asymptotically large $M$, i.e., $\underset{M\rightarrow\infty}{\lim}\frac{C}{\log_2(M)}$, is given in the following proposition.

\underline{\emph{Proposition 2}}: \ If $K=2$, by adopting the passive beamforming structure in \eqref{22} with equal number of elements at the two IRSs ($M_1=M_2=\frac{M}{2}$), the capacity scaling order of an $N_t\times2$ double-IRS aided MIMO system with respect to an asymptotically large $M$ is maximized to $\underset{M\rightarrow\infty}{\lim}\frac{C}{\log_2(M)}=4$.
\begin{IEEEproof}
Please refer to Appendix A.
\end{IEEEproof}

Moreover, in the high transmit power regime, where it is asymptotically optimal to evenly allocate the transmit power to the two eigenchannels ($P_1^\star=P_2^\star=\frac{P}{2}$), \eqref{43} can be well approximated as
\vspace{-0.5em}
\begin{equation}
\vspace{-0.5em}
\begin{aligned}
C&\approx\log_2\bigg(\frac{P^2}{4\sigma^4}(\delta_1\delta_2)^2\bigg)\\
&\overset{\text{(a)}}{=}\log_2\bigg(\frac{P^2}{\sigma^4}N_t^2\Big|a(\mathbf{\Phi}_1)b(\mathbf{\Phi}_2)\Big|^2\big(1-\rho_{N_t}(\Theta_{t})^2\big)\big(1-\rho_{2}(\Theta_{r})^2\big)\bigg),
\label{44}
\end{aligned}
\end{equation}where $\overset{\text{(a)}}{=}$ comes from the expression of $(\delta_1\delta_2)^2$ in \eqref{53}. Surprisingly, the effect of the common phase shifts $\{\gamma_1,\gamma_2\}$ and the double-reflection link $\check{c}(\mathbf{\Phi}_1,\mathbf{\Phi}_2)$ is negligible. This is because the spatial multiplexing gain introduced by the two single-reflection links $a(\mathbf{\Phi}_1)$ and $b(\mathbf{\Phi}_2)$ now takes the dominant effect on the MIMO channel capacity. In the high transmit power regime, it can be shown that the passive beamforming structure in \eqref{22} maximizes the channel capacity in \eqref{44} by maximizing $\big|a(\mathbf{\Phi}_1)b(\mathbf{\Phi}_2)\big|=\big|\frac{\alpha^2M_1M_2}{d_{R_1}\!d_{T_1}\!d_{R_2}\!d_{T_2}}\big|$, and it is optimal to further allocate even number of elements at the two IRSs ($M_1=M_2=\frac{M}{2}$). Thus, we have the following proposition.

\underline{\emph{Proposition 3}}: \ If $K=2$, by adopting the passive beamforming structure in \eqref{22} with equal number of elements at the two IRSs ($M_1=M_2=\frac{M}{2}$), the capacity of an $N_t\times 2$ double-IRS aided MIMO system with asymptotically large $P$ is maximized as
\vspace{-0.2em}
\begin{equation}
\vspace{-0.2em}
\begin{aligned}
C=\log_2\bigg(\frac{\alpha^4N_t^2M^4P^2}{16d_{R_1}^2\!d_{T_1}^2\!d_{R_2}^2\!d_{T_2}^2\sigma^4}\big(1-\rho_{N_t}(\Theta_{t})^2\big)\big(1-\rho_{2}(\Theta_{r})^2\big)\bigg).
\label{45}
\end{aligned}
\end{equation}

From \eqref{45} we can see that the capacity scaling order with respect to an asymptotically large $P$ is $\underset{P\rightarrow\infty}{\lim}\frac{C}{\log_2(P)}=2$, which means that a $2$-fold spatial multiplexing gain can be achieved in the high transmit power regime. It is also worth noting that when $\rho_{N_t}(\Theta_{t})=0$ (i.e., orthogonal $\mathbf{t}_{1R}$ and $\mathbf{t}_{2R}$) and $\rho_{2}(\Theta_{r})=0$ (orthogonal $\mathbf{r}_{1L}$ and $\mathbf{r}_{2L}$), the MIMO channel capacity in \eqref{45} is maximized.

\begin{figure}[t]
\centering
\includegraphics[width=5.1in]{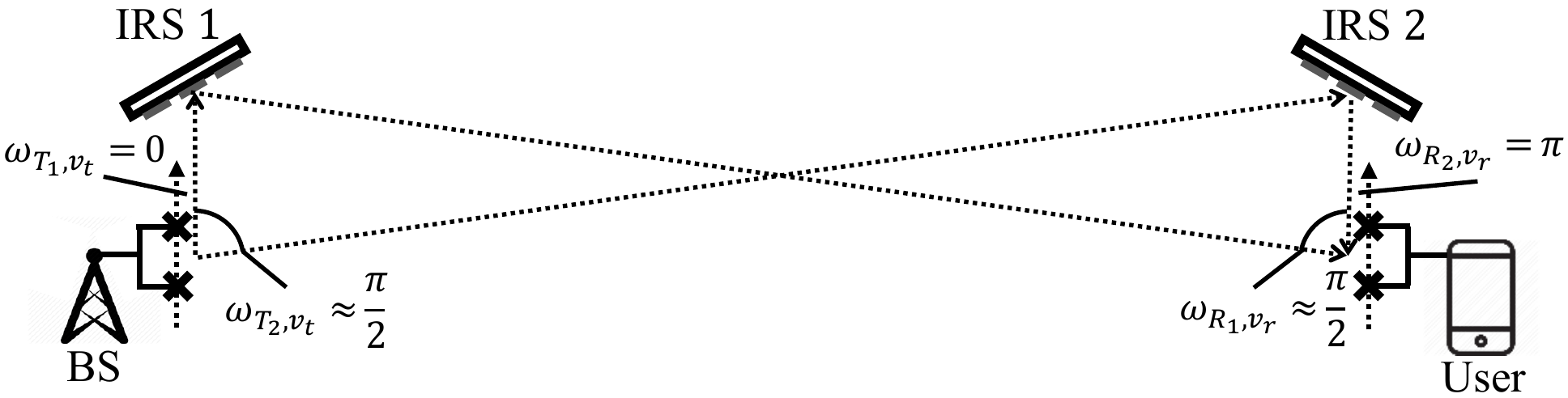}
\caption{Rank-two $\mathbf{H}$ with orthogonal $\mathbf{t}_{1R}$ and $\mathbf{t}_{2R}$ as well as orthogonal $\mathbf{r}_{1L}$ and $\mathbf{r}_{2L}$, under $N_t=N_r=2$ and $l_n=\frac{\lambda}{2}$.}
\vspace{-1.5em}
\label{geo1}
\end{figure}

In Fig.~\ref{geo1}, we illustrate a scenario with orthogonal $\mathbf{t}_{1R}$ and $\mathbf{t}_{2R}$ as well as orthogonal $\mathbf{r}_{1L}$ and $\mathbf{r}_{2L}$, under $N_t=N_r=2$ and $l_n=\frac{\lambda}{2}$. The two IRSs are deployed such that $\omega_{T_1,v_t}=0$ and $\omega_{T_2,v_t}\approx\frac{\pi}{2}$, also $\omega_{R_1,v_r}\approx\frac{\pi}{2}$ and $\omega_{R_2,v_r}=\pi$, thus $\mathbf{H}$ is of rank-two due to $\omega_{T_1,v_t}\neq\omega_{T_2,v_t}$ and $\omega_{R_1,v_r}\neq\omega_{R_2,v_r}$ according to \eqref{42}. We further have $\Theta_{t}=\frac{l_{n}}{\lambda}\big(\cos(\omega_{T_2,v_t})-\cos(\omega_{T_1,v_t})\big)=-\frac{1}{2}$ and $\Theta_{r}=\frac{l_{n}}{\lambda}\big(\cos(\omega_{R_1,v_r})-\cos(\omega_{R_2,v_r})\big)=\frac{1}{2}$, thus $\rho_{2}(\Theta_{t})=0$ and $\rho_{2}(\Theta_{r})=0$, i.e., orthogonal $\mathbf{t}_{1R}$ and $\mathbf{t}_{2R}$ as well as orthogonal $\mathbf{r}_{1L}$ and $\mathbf{r}_{2L}$. 

\subsection{Capacity Scaling with Rank-one $\mathbf{H}$}

We now consider the case that $\mathbf{H}$ in \eqref{21} is of rank-one. Here, we assume $\rho_{N_t}(\Theta_{t})=1$, i.e., $\mathbf{t}_{1R}=\mathbf{t}_{2R}$, for the purpose of exposition,\footnote{All the analysis is directly applicable to the other case of $\mathbf{r}_{1L}=\mathbf{r}_{2L}$.} $\mathbf{H}$ can be thus rewritten as
\vspace{-0.5em}
\begin{equation}
\vspace{-0.5em}
\begin{aligned}
\mathbf{H}=\Big(a(\mathbf{\Phi}_1)\mathbf{r}_{1L}+\big(b(\mathbf{\Phi}_2)+\check{c}(\mathbf{\Phi}_1,\mathbf{\Phi}_2)\big)\mathbf{r}_{2L}\Big)\mathbf{t}_{1R}^T,
\label{46}
\end{aligned}
\end{equation}which yields only one singular value $\delta_1$, and the MIMO channel capacity can be expressed as
\vspace{-0.5em}
\begin{equation}
\vspace{-0.5em}
\begin{aligned}
C=\log_2\bigg(1+\frac{P\delta_1^2}{\sigma^2}\bigg),
\label{47}
\end{aligned}
\end{equation}with its exact expression given in the following proposition.

\underline{\emph{Proposition 4}}: \ If $K=1$, by adopting the passive beamforming structure in \eqref{22} and setting the common phase shifts as $\gamma_1^\star=\frac{\beta_b}{\beta_c}$ and $\gamma_2^\star=\frac{1}{\beta_b}\big(\frac{\beta_a\beta_b}{\beta_c}-\pi\Theta_r\big)$, the capacity of an $N_t\times2$ double-IRS aided MIMO system is maximized as
\vspace{-0.5em}
\begin{equation}
\vspace{-0.5em}
\begin{aligned}
C\!\!=\!\log_2\!\!\Bigg(\!1\!\!+\!\frac{2N_tP}{\sigma^2}\!\bigg(\!\!\Big(\frac{\alpha M_1}{d_{R_1}\!d_{T_1}}\!\Big)^2\!\!\!+\!\Big(\frac{\alpha M_2}{d_{R_2}\!d_{T_2}}\!+\!\frac{\alpha^{3/2} M_1\!M_2}{d_{R_2}\!d_{S}d_{T_1}}\!\Big)^2\!\!\!+\!2\Big(\frac{\alpha M_1}{d_{R_1}\!d_{T_1}}\!\Big)\Big(\frac{\alpha M_2}{d_{R_2}\!d_{T_2}}\!+\!\frac{\alpha^{3/2} M_1\!M_2}{d_{R_2}\!d_{S}d_{T_1}}\!\Big)\rho_{2}(\Theta_r)\!\bigg)\!\!\Bigg).
\label{48}
\end{aligned}
\end{equation}
\begin{IEEEproof}
Please refer to Appendix B.
\end{IEEEproof}

Note that for large $M$, it is asymptotically optimal to allocate even number of elements at the two IRSs ($M_1=M_2=\frac{M}{2}$), since $\frac{\alpha^{3/2} M_1M_2}{d_{R_2}\!d_{S}d_{T_1}}$ will be much larger than $\frac{\alpha M_1}{d_{R_1}\!d_{T_1}}$ or $\frac{\alpha M_2}{d_{R_2}\!d_{T_2}}$, and from \eqref{48} we have
\vspace{-0.5em}
\begin{equation}
\vspace{-0.5em}
\begin{aligned}
C&=\log_2\Bigg(1+\frac{2N_tM^4P}{\sigma^2}\bigg(\Big(\frac{\alpha}{2Md_{R_1}\!d_{T_1}}\Big)^2\!+\!\Big(\frac{\alpha}{2Md_{R_2}\!d_{T_2}}\!+\!\frac{\alpha^{3/2} }{4d_{R_2}\!d_{S}d_{T_1}}\Big)^2+\frac{\alpha}{Md_{R_1}\!d_{T_1}}\\
&\ \ \ \ \ \ \ \ \ \ \ \ \ \ \ \ \ \ \ \ \ \ \ \ \ \ \ \ \ \ \ \ \ \ \ \ \times\Big(\frac{\alpha}{2Md_{R_2}\!d_{T_2}}\!+\!\frac{\alpha^{3/2}}{4d_{R_2}\!d_{S}d_{T_1}}\Big)\rho_{2}(\Theta_r)\bigg)\Bigg).
\label{49}
\end{aligned}
\end{equation}As we can see from \eqref{49}, although the spatial multiplexing gain is lost, i.e.,$\underset{P\rightarrow\infty}{\lim}\frac{C}{\log_2(P)}=1$, due to the rank-one $\mathbf{H}$, a capacity scaling order of $4$ with respect to an asymptotically large $M$ can still be achieved, i.e., $\underset{M\rightarrow\infty}{\lim}\frac{C}{\log_2(M)}=4$, thanks to the cooperative passive beamforming gain of the double-reflection link. It is worth noting that for the rank-one $\mathbf{H}$ with $\rho_{2}(\Theta_{t})=1$ (i.e., $\mathbf{t}_{1R}=\mathbf{t}_{2R}$), the MIMO channel capacity in \eqref{49} is maximized with $\rho_{2}(\Theta_{r})=1$ ($\mathbf{r}_{1L}=\mathbf{r}_{2L}$).

\begin{figure}[t]
\centering
\includegraphics[width=4.3in]{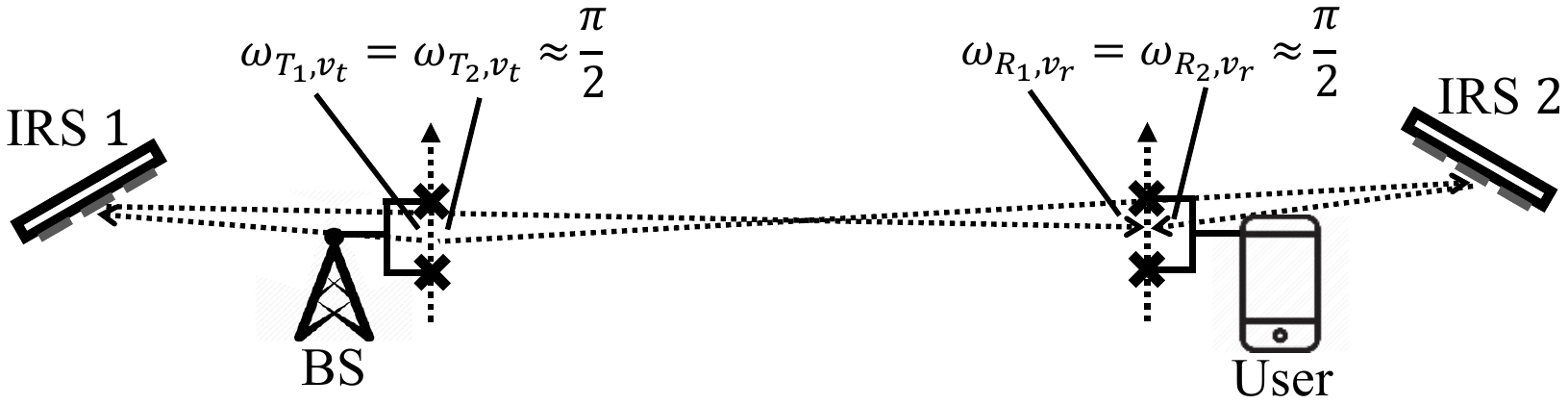}
\caption{Rank-one $\mathbf{H}$ with $\mathbf{t}_{1R}=\mathbf{t}_{2R}$ and $\mathbf{r}_{1L}=\mathbf{r}_{2L}$, under $N_t=N_r=2$ and $l_n=\frac{\lambda}{2}$.}
\vspace{-1.5em}
\label{geo2}
\end{figure}

In Fig.~\ref{geo2}, we visualize a possible scenario with $\mathbf{t}_{1R}=\mathbf{t}_{2R}$ as well as $\mathbf{r}_{1L}=\mathbf{r}_{2L}$, under $N_t=N_r=2$ and $l_n=\frac{\lambda}{2}$. The two IRSs are deployed such that $\omega_{T_1,v_t}=\omega_{T_2,v_t}=\omega_{R_1,v_r}=\omega_{R_2,v_r}\approx\frac{\pi}{2}$, thus we have $\rho_{2}(\Theta_{t})=1$ and $\rho_{2}(\Theta_{r})=1$, and consequently $\mathbf{H}$ is of rank-one.

\subsection{Comparison with Benchmark Single-IRS System}

For comparison with our considered double-IRS system, we also present the conventional single-IRS aided MIMO system \cite{zhang2020capacity}, where only IRS $2$ is deployed with all $M$ reflecting elements, thus only one single-reflection link is available. The effective MIMO channel is of rank-one and given by
\vspace{-0.5em}
\begin{equation}
\vspace{-0.5em}
\begin{aligned}
\mathbf{H}=\mathbf{R}\mathbf{\Phi}\mathbf{T}=\frac{\alpha\beta_b}{d_{R_2}d_{T_2}}\mathbf{r}_{L}\mathbf{r}_{R}^T\mathbf{\Phi}\mathbf{t}_{L}\mathbf{t}_{R}^T,
\label{50}
\end{aligned}
\end{equation}where $\mathbf{T}\in\mathbb{C}^{M\times N_t}$ and $\mathbf{R}\in\mathbb{C}^{2\times M}$ are the channels from the BS to the single IRS and from the single IRS to the user, respectively, following similar definitions as in \eqref{10} and \eqref{13}, and $\mathbf{\Phi}=\diag\{\phi_{1},\cdots,\phi_{M}\}\in\mathbb{C}^{M\times M}$ is the passive beamforming matrix of the single IRS.

It can be easily shown that by deploying the optimal passive beamformer $\phi_{m}^{\star}=(\mathbf{r}_{R})_{m}^*(\mathbf{t}_{L})_{m}^*$, $m\in\mathcal{M}$, the MIMO channel capacity is maximized as
\vspace{-0.5em}
\begin{equation}
\vspace{-0.5em}
\begin{aligned}
C=\log_2\bigg(1+\frac{2\alpha^2N_tM^2P}{d_{R_2}^2d_{T_2}^2\sigma^2}\bigg).
\label{51}
\end{aligned}
\end{equation}As we can see from \eqref{51}, there is neither spatial multiplexing gain harvested from the two single-reflection links, nor cooperative passive beamforming gain brought by the double-reflection link, as in the double-IRS aided MIMO system. Thus, the capacity scaling order with respect to $M$ and $P$ are $\underset{M\rightarrow\infty}{\lim}\frac{C}{\log_2(M)}=2$ and $\underset{P\rightarrow\infty}{\lim}\frac{C}{\log_2(P)}=1$, respectively, for the single-IRS aided MIMO system.

\underline{\emph{Theroem 1}}: \ The capacity scaling orders for the $N_t\times2$ MIMO system aided by double IRSs and single IRS, with respect to asymptotically large $M$ or $P$, are given in Table II.
\begin{table}[t]
\centering
\caption{Capacity scaling orders under LoS channels for $N_t\times2$ MIMO system}
\begin{tabular}{|c|c|l|c|c|}
\hline
\multicolumn{3}{|c|}{\multirow{2}{*}{System}}                    & \multicolumn{2}{c|}{Capacity scaling order} \\ \cline{4-5} 
\multicolumn{3}{|c|}{}                                           & $\underset{P\rightarrow\infty}{\lim}\frac{C}{\log_2(P)}$   & $\underset{M\rightarrow\infty}{\lim}\frac{C}{\log_2(M)}$   \\ \hline
\multicolumn{3}{|c|}{Single IRS}                                    & 1                    & 2                    \\ \hline
\multirow{2}{*}{Double IRSs}      & \multicolumn{2}{c|}{Rank-one $\mathbf{H}$} & 1                    & 4                    \\ \cline{2-5} 
                               & \multicolumn{2}{c|}{Rank-two $\mathbf{H}$} & 2                    & 4                    \\ \hline
\end{tabular}
\vspace{-1.5em}
\end{table}

As we can see from Table II, the capacity scaling order of our considered double-IRS system with respect to an asymptotically large $M$ is $\underset{M\rightarrow\infty}{\lim}\frac{C}{\log_2(M)}=4$, regardless of the rank of $\mathbf{H}$, which substantially outperforms the conventional single-IRS system. While by further deploying the IRS locations as in \eqref{41}, we can potentially harvest a $2$-fold spatial multiplexing gain from the double-IRS system, i.e., $\underset{P\rightarrow\infty}{\lim}\frac{C}{\log_2(P)}=2$.

\begin{spacing}{1.49}

\section{Numerical Results}

\begin{figure}[t]
\centering
\includegraphics[width=4.0in]{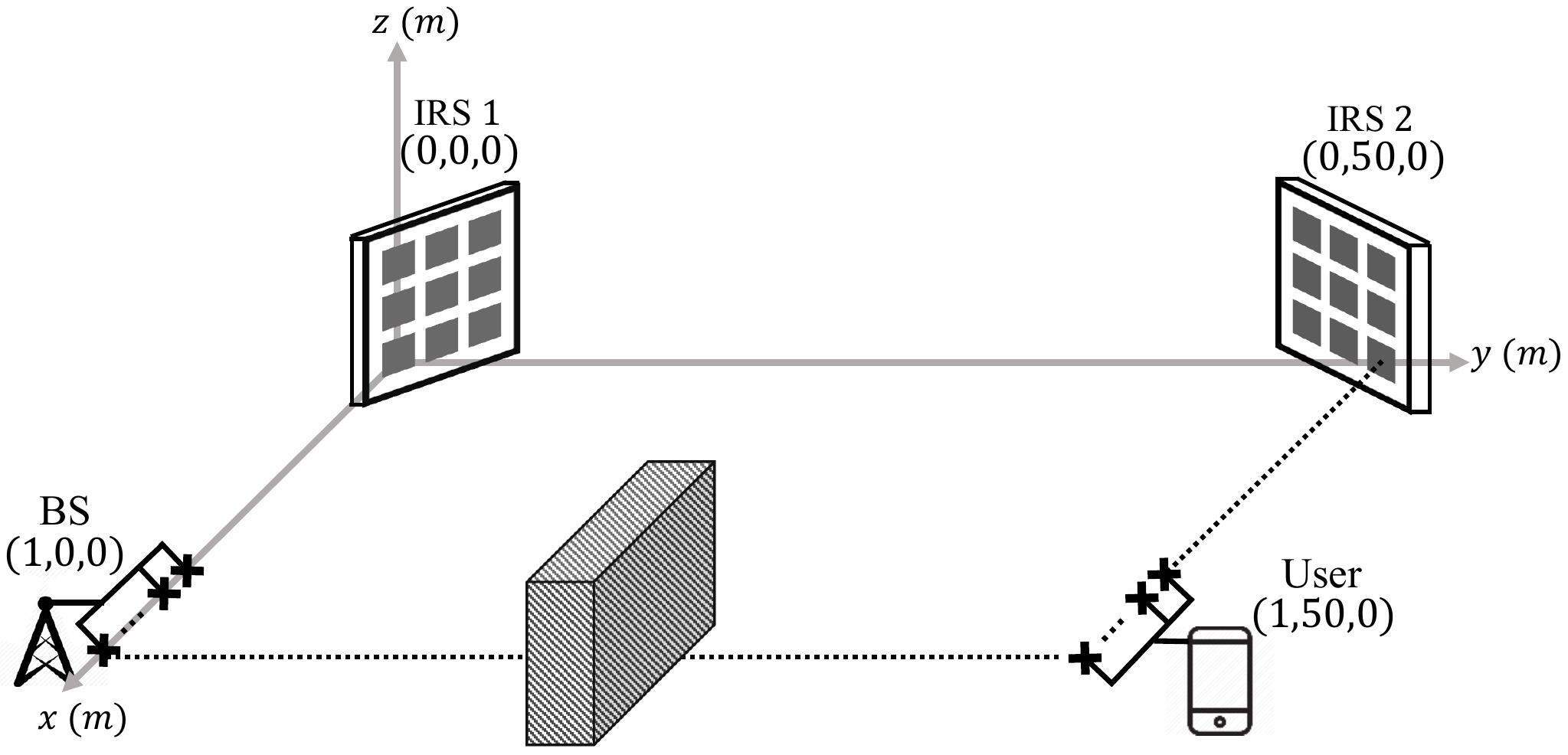}
\caption{Illustration of the simulation setup.}
\vspace{-1.8em}
\label{simu3d}
\end{figure}

In this section, we provide extensive numerical results to evaluate the performance of our proposed Algorithm 1 in Section IV and validate our capacity scaling analysis in Section V. We set the 3D spatial locations of the BS, user, IRS 1, and IRS 2 as $\mathbf{u}_t=[1,0,0]^T$, $\mathbf{u}_r=[1,50,0]^T$, $\mathbf{u}_1=[0,0,0]^T$, and $\mathbf{u}_2=[0,50,0]^T$, respectively, as illustrated in Fig.~\ref{simu3d}. We set ULA at the BS and user, with the base direction of BS's antenna array and that of user's antenna array both set as $\mathbf{v}_t=\mathbf{v}_r=[-1,0,0]^T$. We set URA at IRS $1$ and IRS $2$, with the base directions of IRS $1$ set as $\mathbf{v}_1^{(a)}=[-\frac{1}{2},\frac{\sqrt{3}}{2},0]^T$ and $\mathbf{v}_1^{(b)}=[0,0,1]^T$, while those of IRS $2$ set as $\mathbf{v}_2^{(a)}=[0,0,1]^T$ and $\mathbf{v}_2^{(b)}=[-\frac{1}{2},-\frac{\sqrt{3}}{2},0]^T$. The carrier frequency is set as $3.5$ GHz, thus the wavelength is $\lambda=0.087$ m and the channel power gain at reference distance of $1$ m is $\alpha=-43$ dB \cite{goldsmith2005wireless}. The antenna spacing is set as $l_{n}=\frac{\lambda}{2}$, while the IRS element spacing is set as $l_{m}=\frac{\lambda}{10}$. Note that under the above setup, we have $\omega_{T_1,v_t}\neq\omega_{T_2,v_t}$ as well as $\omega_{R_1,v_r}\neq\omega_{R_2,v_r}$, and the effective MIMO channel $\mathbf{H}$ is of rank-two according to \eqref{42}. The noise power level at the user receiver is set as $\sigma^2 = -70$ dBm. All the involved channels are LoS following the definitions in \eqref{2}. For all the double-IRS aided MIMO systems considered in this section, we set $M_1=M_2=\frac{M}{2}$ since it maximizes the power gain of the double-reflection link. The convergence threshold of Algorithm 1 in terms of the relative increment in (P2)'s objective function is set as $\epsilon=10^{-5}$.

\subsection{Comparison of Algorithm 1 with Other Capacity Maximization Algorithms for Double-IRS Aided MIMO System}

\begin{figure}[t]
\centering
\subfigure[Low transmit power regime with $P=-10$ dBm.]{
\includegraphics[width=2.0in]{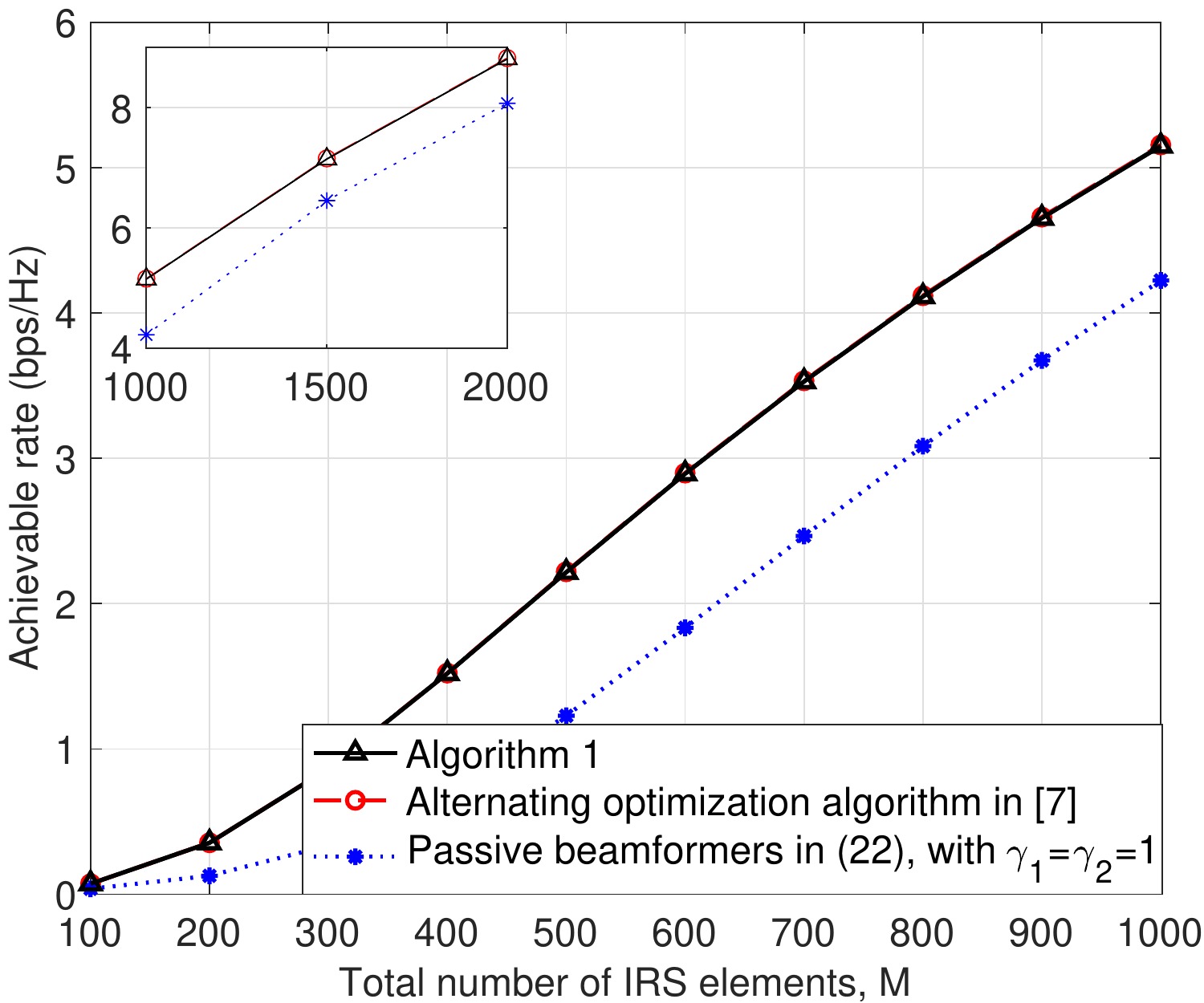}}
\hspace{0.01in}
\subfigure[High transmit power regime with $P=20$ dBm.]{
\includegraphics[width=2.0in]{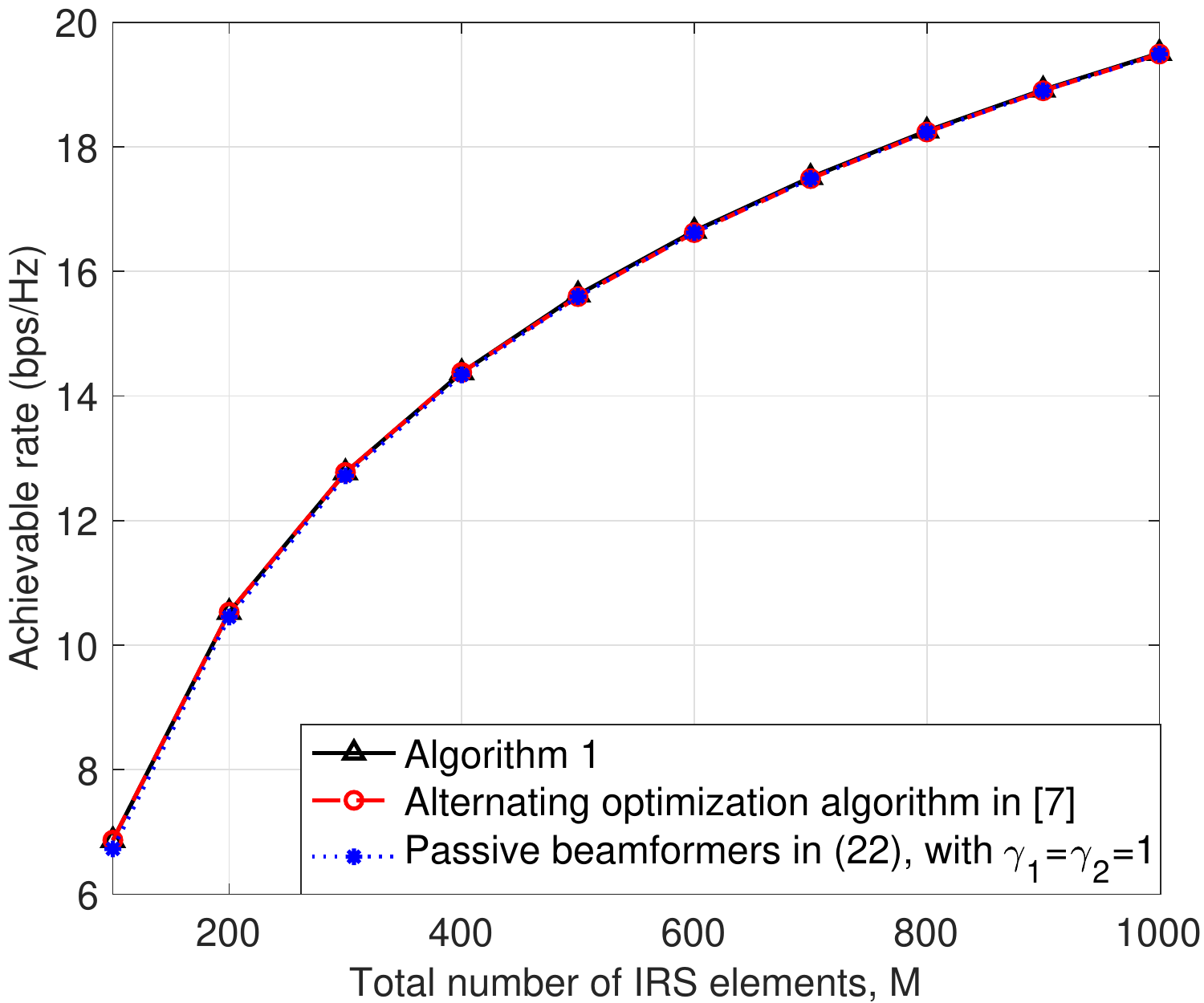}}
\hspace{0.01in}
\subfigure[Computational time versus total number of IRS elements.]{
\includegraphics[width=2.0in]{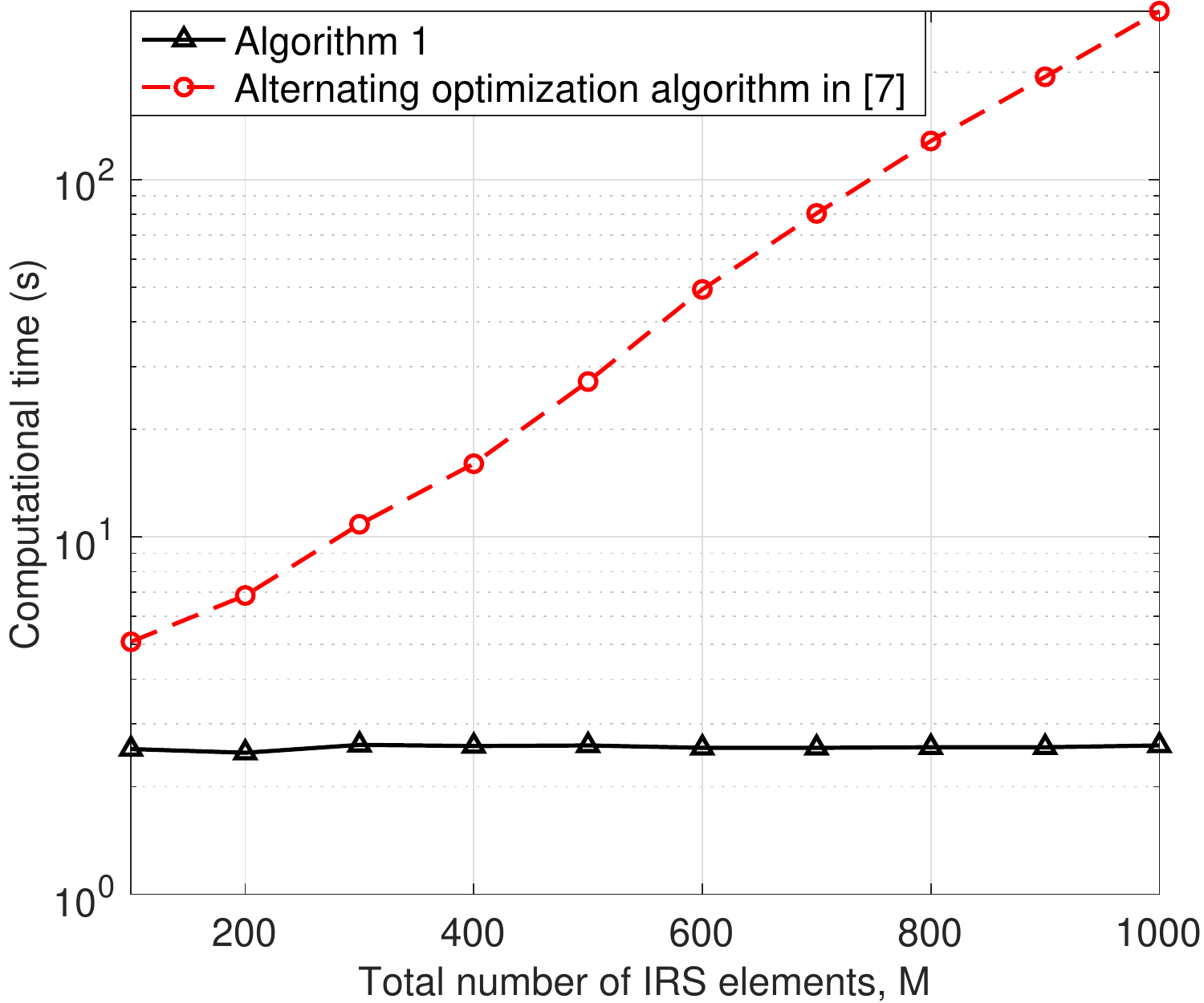}}
\caption{Comparison of different capacity maximization algorithms.}
\vspace{-1.8em}
\label{alg}
\end{figure}

First, we show the advantages of our proposed Algorithm 1, by comparing with two benchmark algorithms:
\begin{itemize}
\item{The alternating optimization algorithm based on \cite{zhang2020capacity}, by iteratively optimizing one variable in $\mathbf{Q}\cup\{\phi_{1,m_1}\}_{m_1\in\mathcal{M}_1}\cup\{\phi_{2,m_2}\}_{m_2\in\mathcal{M}_2}$ at each time with the other $M$ variables being fixed.}
\item{A heuristic design by directly deploying the passive beamformers in \eqref{22} with $\gamma_1=\gamma_2=1$, i.e., without optimizing the common phase shifts of the two IRSs.}
\end{itemize}
We consider the setup with $N_t=N_r=3$ and evaluate the achievable rate in bps/Hz by all the three algorithms for both the low transmit power regime with $P=-10$ dBm in Fig.~\ref{alg}(a) and the high transmit power regime with $P=20$ dBm in Fig.~\ref{alg}(b). In Fig.~\ref{alg}(c), we plot the computational time in second (s) versus the total number of IRS elements $M$, using Algorithm 1 and the alternating optimization algorithm based on \cite{zhang2020capacity}, respectively.

It can be observed from Fig.~\ref{alg}(a) and Fig.~\ref{alg}(b) that for both power regimes, our proposed low-complexity Algorithm 1 can achieve almost the same performance as the alternating optimization algorithm based on \cite{zhang2020capacity}. Note that the computational time of the former is independent with $M$ thanks to the exploitation of the LoS channels, while that of the latter increases fast with $M$.

Moreover, in the low transmit power regime (see Fig.~\ref{alg}(a)), our proposed algorithm significantly outperforms the heuristic design without optimizing the common phase shifts, e.g., by $25\%$ at $M=1000$. This is because in the low transmit power regime, it is optimal to transmit a single data stream, thus the achievable rate critically depends on the power gain of the strongest ($1$st) eigenchannel, and the maximization of which requires a fine-tuning of the common phase shifts such that the three reflection links can be coherently combined. On the other hand, in the high transmit power regime (see Fig.~\ref{alg}(b)), the heuristic design performs closely to our proposed algorithm as well as the alternating optimization algorithm based on \cite{zhang2020capacity}. This is because the achievable rate in the high transmit power regime is dominated by the spatial multiplexing (rank) gain, and all the three algorithms achieve the maximum rank of $2$ with the considered IRS deployment. Furthermore, it can be observed that for both the low and high transmit power regimes, the achievable rate increases by $4$ bps/Hz when doubling $M$. This shows that the capacity scaling order with respect to an asymptotically large $M$ for $N_t=N_r=3$ is still $4$, which is the same as our analytical results for $N_t\times2$ MIMO system in Table II.

\subsection{Double-IRS Aided Versus Single-IRS Aided MIMO Systems}

\begin{figure*}[t]
\centering
\subfigure[Achievable rate versus total number of IRS elements in the low transmit power regime with $P=-10$ dBm.]{
\includegraphics[width=3.0in]{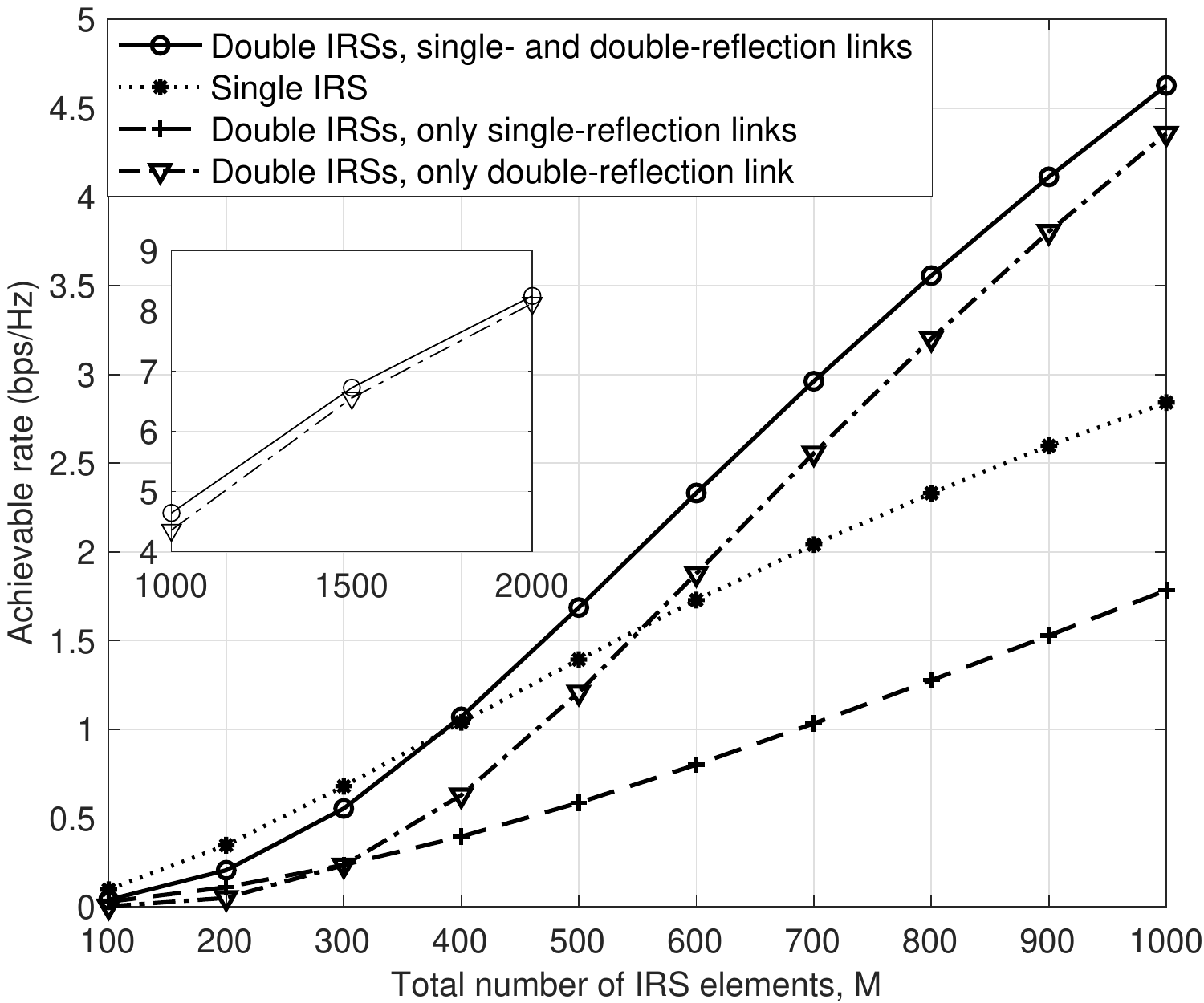}}
\hspace{0.01in}
\subfigure[Achievable rate versus total number of IRS elements in the moderate transmit power regime with $P=5$ dBm.]{
\includegraphics[width=3.0in]{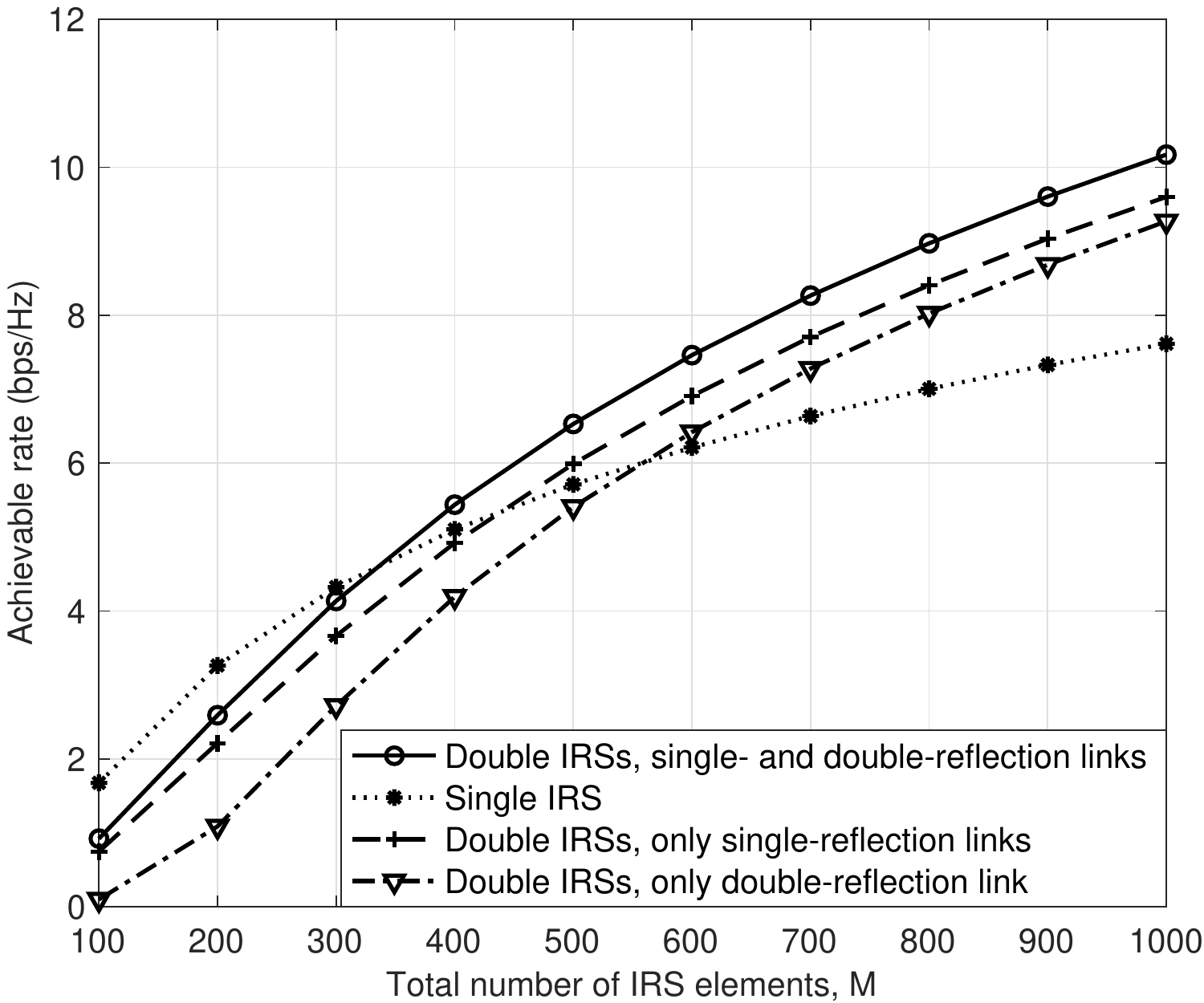}}
\hspace{0.01in}
\subfigure[Achievable rate versus total number of IRS elements in the high transmit power regime with $P=20$ dBm.]{
\includegraphics[width=3.0in]{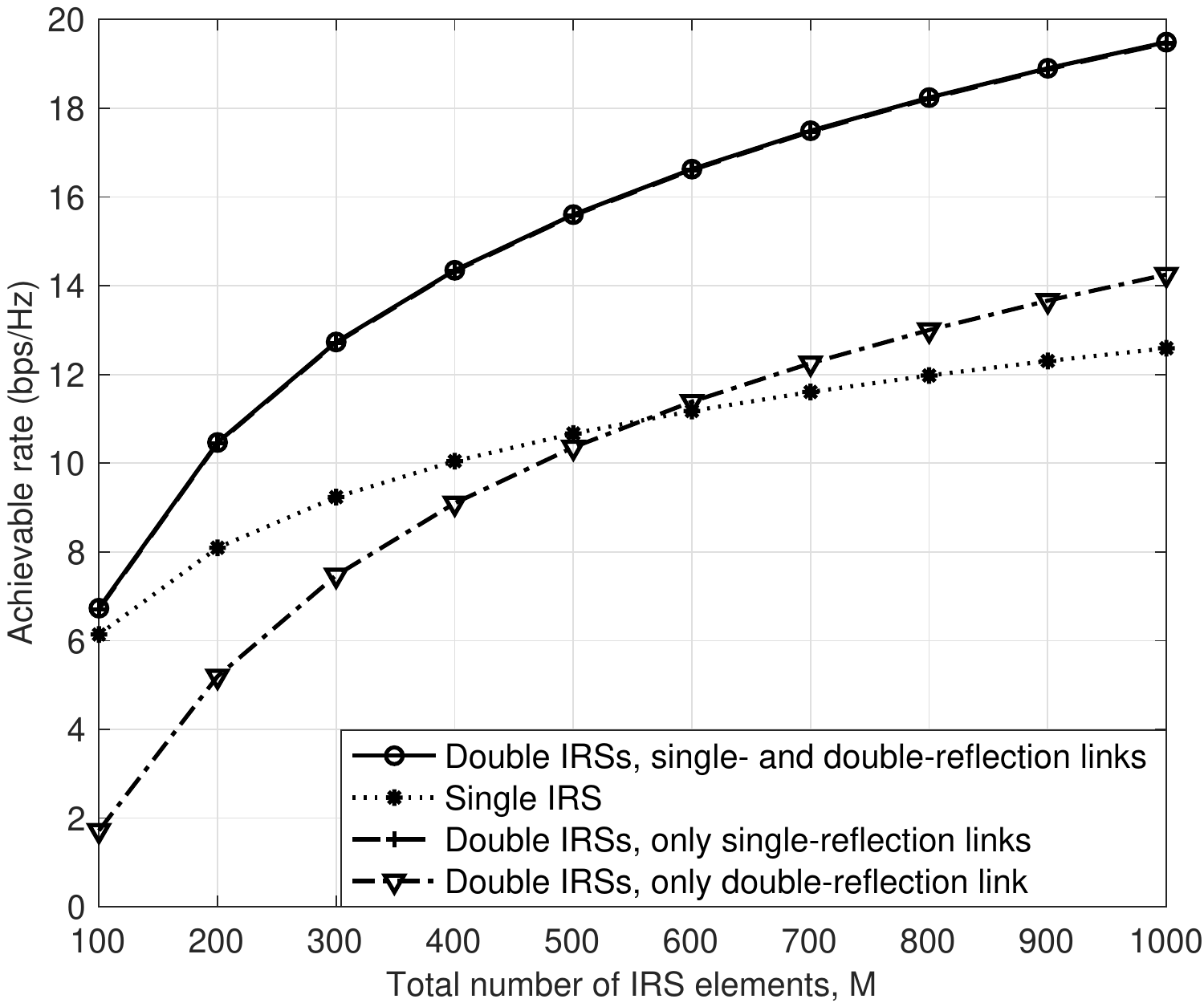}}
\hspace{0.01in}
\subfigure[Achievable rate versus transmit power with $M=1000$.]{
\includegraphics[width=3.0in]{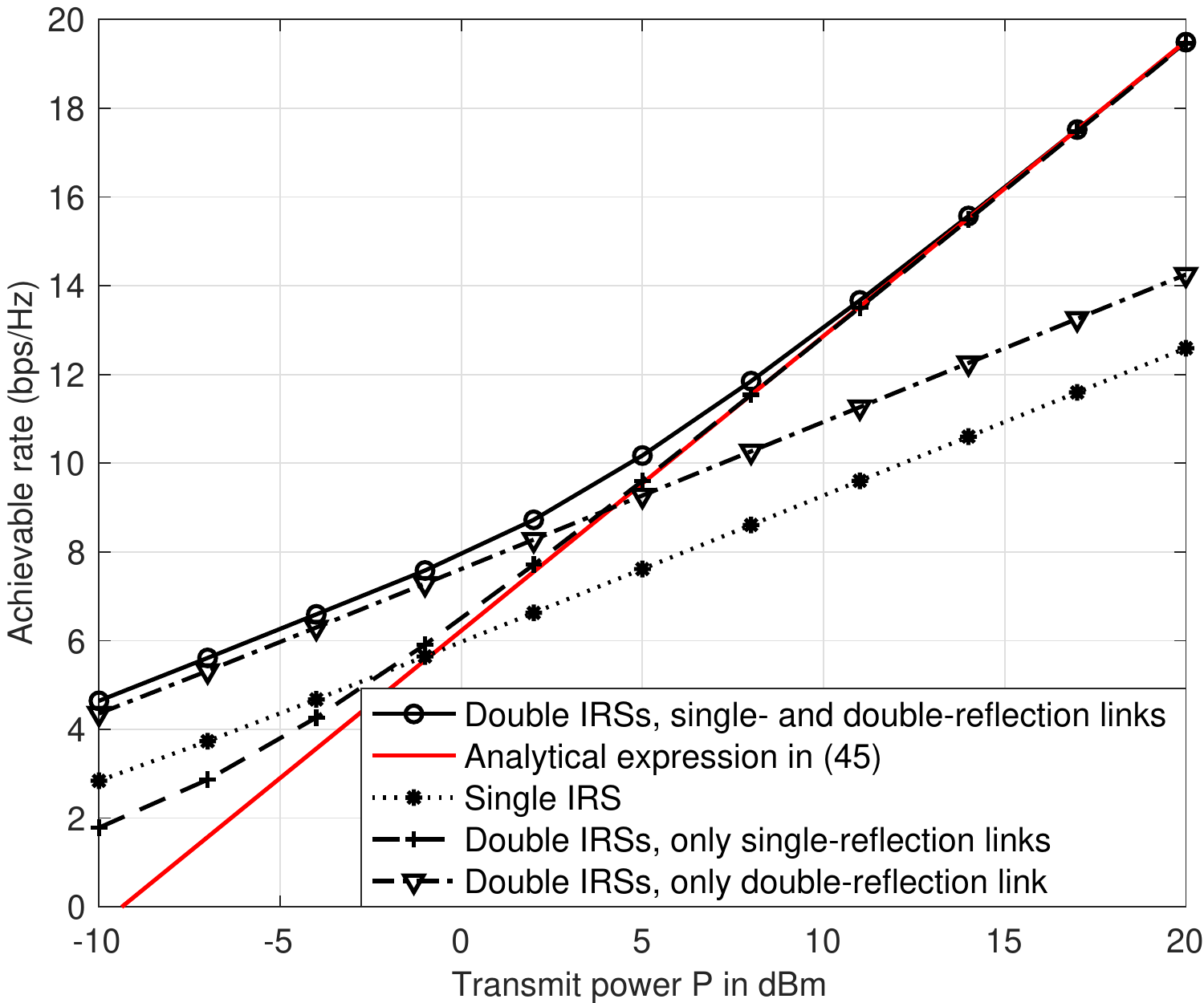}}
\caption{Performance comparison between our proposed double-IRS system and the conventional single-IRS system.}
\vspace{-1.8em}
\label{cvm}
\end{figure*}

Next, we compare in Fig.~\ref{cvm} the performance of the double-IRS aided and single-IRS aided MIMO systems by considering a $4\times2$ MIMO system, for which the capacity scaling orders of both systems are derived in Section V. For the double-IRS aided MIMO system, we use our proposed Algorithm 1 for the capacity maximization. For the single-IRS aided MIMO system, we consider that all the $M$ elements are equipped at IRS 2 (following a similar setup as in Section V-D). To identify which of the single-reflection and double-reflection links is performance-dominant in different transmit power regimes, we consider two benchmark scenarios for the double-IRS system, i.e., when only the two single-reflection links are present, and when only the double-reflection link is present, respectively. 

In Figs.~\ref{cvm}(a)--(c), we plot the achievable rate versus the total number of IRS elements $M$ for different transmit power regimes. In addition, we show in Fig.~\ref{cvm}(d) the achievable rate versus the transmit power $P$ with $M=1000$. It is observed from Figs.~\ref{cvm}(a)--(d) that the double-IRS system achieves higher achievable rate over its single-IRS counterpart as long as the number of IRS elements $M$ or the transmit power $P$ is not small. The capacity scaling orders with respect to asymptotically large $M$ or $P$ for both double-IRS and single-IRS systems derived in Table II are also verified.

Moreover, for the low transmit power regime with $P=-10$ dBm in Fig.~\ref{cvm}(a), it can be observed that the double-IRS system with only the double-reflection link performs closely to that with both single-reflection and double-reflection links. This is because the achievable rate in the low transmit power regime is maximized by transmitting a single data stream over the strongest eigenchannel, i.e., the \emph{beamforming mode}, thus the cooperative passive beamforming gain harvested from the double-reflection link plays a dominant role. On the other hand, for the high transmit power regime with $P=20$ dBm in Fig.~\ref{cvm}(c), the double-IRS system with only the single-reflection links performs similarly as that with both links, since the MIMO system now operates in the \emph{spatial multiplexing mode} and the rank gain achieved by the two single-reflection links is dominant. While for the moderate transmit power regime with $P=5$ dBm in Fig.~\ref{cvm}(b), both the single-reflection and double-reflection links contribute to the overall achievable rate of the double-IRS system in general.

Furthermore, it can be observed from Fig.~\ref{cvm}(d) that as the transmit power $P$ increases, the main contributor of the double-IRS system's achievable rate changes from the double-reflection link's cooperative passive beamforming gain to the two single-reflection links' spatial multiplexing gain. We also plot the analytical expression of the MIMO channel capacity with asymptotically large $P$ for rank-two $\mathbf{H}$ in \eqref{45}, which coincides with the achievable rate by employing our proposed Algorithm 1 in the high transmit power regime. This validates our analytical results and also shows that our proposed Algorithm 1 can achieve a near-optimal solution to (P1).

\begin{figure*}[t]
\centering
\subfigure[Singular values of $\mathbf{H}$ (after alternating optimization of $\{\mathbf{Q},\gamma_1$,$\gamma_2\}$) versus total number of IRS elements.]{
\includegraphics[width=3.0in]{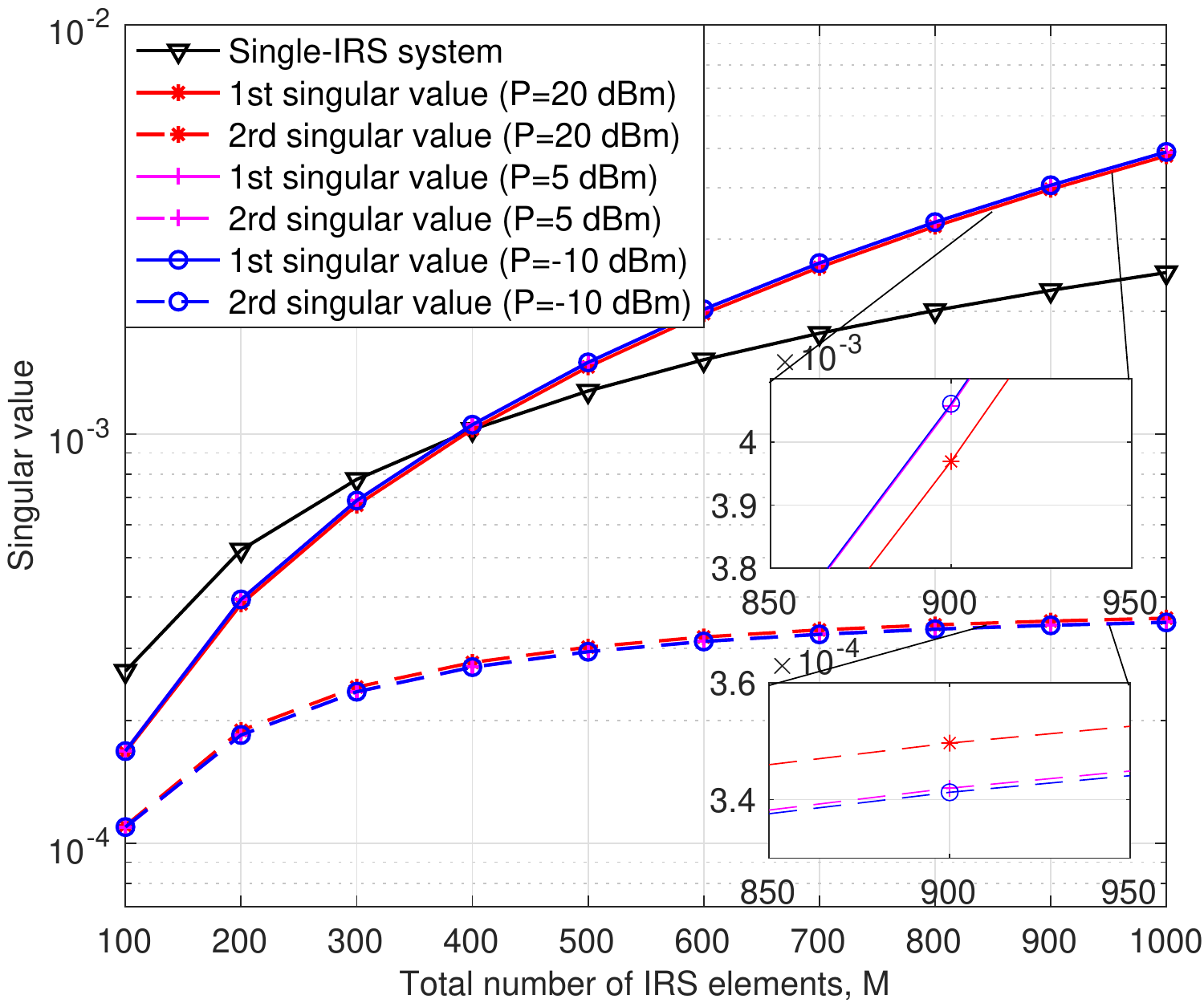}}
\hspace{0.01in}
\subfigure[Power allocation versus total number of IRS elements.]{
\includegraphics[width=3.0in]{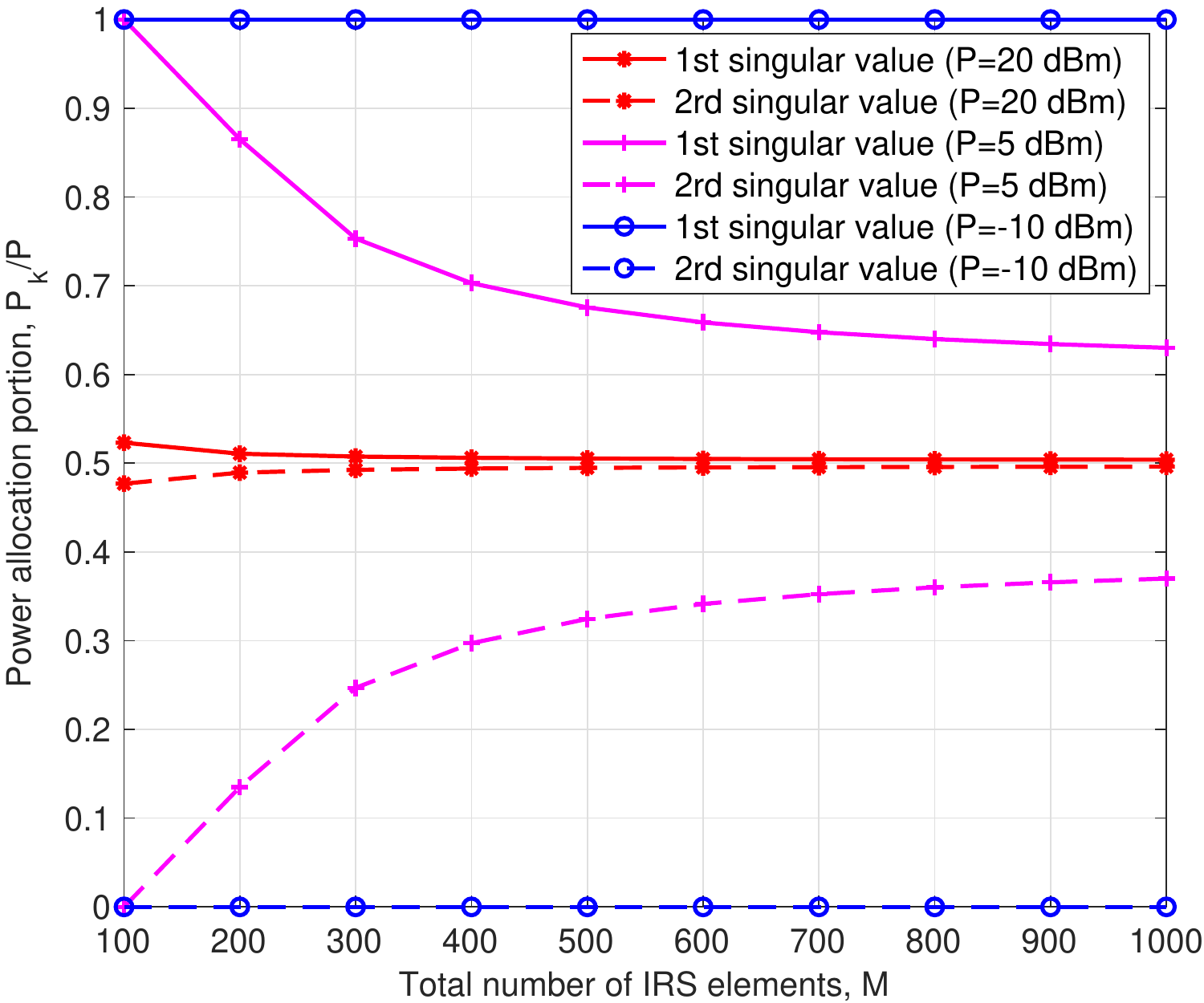}}
\caption{Singular values of double-IRS aided MIMO channel and corresponding transmit power allocation.}
\vspace{-1.8em}
\label{other}
\end{figure*}

In addition, we show in Fig.~\ref{other} the singular values of the effective MIMO channel $\mathbf{H}$ as well as the corresponding transmit power allocation for the double-IRS system in different transmit power regimes. It can be observed from Fig.~\ref{other}(a) that the singular values of $\mathbf{H}$ vary with different $P$'s since the IRSs' common phase shifts are iteratively optimized with the transmit covariance matrix (the power allocation over the different eigenchannels). Particularly, for the small-to-moderate transmit power regime, e.g., $P=-10$ dBm and $P=5$ dBm, it can be observed from Fig.~\ref{other}(b) that more power needs to be allocated to the strongest eigenchannel, and consequently the strongest ($1$st) singular value is adjusted to be larger comparing with that in the high transmit power regime, e.g., $P=20$ dBm, as illustrated in Fig.~\ref{other}(a).

On the other hand, for the high transmit power regime, e.g., $P=20$ dBm, it is asymptotically optimal to evenly allocate the transmit power on the two eigenchannels as shown in Fig.~\ref{other}(b), thus our proposed algorithm will strike a balance between the two singular values as shown in Fig.~\ref{other}(a). The above results show that our proposed algorithm for determining the transmit covariance matrix and IRS common phase shifts is able to adaptively tune the effective MIMO channel $\mathbf{H}$ according to the available transmit power $P$. For comparison, we also show the sole singular value of the effective MIMO channel under the single-IRS system, which is observed to be larger than the strongest singular value under the double-IRS system when $M$ is small, and becomes smaller than the latter as $M$ increases. This is because as $M$ increases, the cooperative passive beamforming gain of the double-IRS system increases much faster than the passive beamforming gain of the single-IRS system ($\mathcal{O}(M^4)$ versus $\mathcal{O}(M^2)$).

\subsection{Performance under Different User-IRS Angles}
\begin{figure*}[t]
\subfigure[A top-down view of the location setup.]{
\includegraphics[width=3.0in]{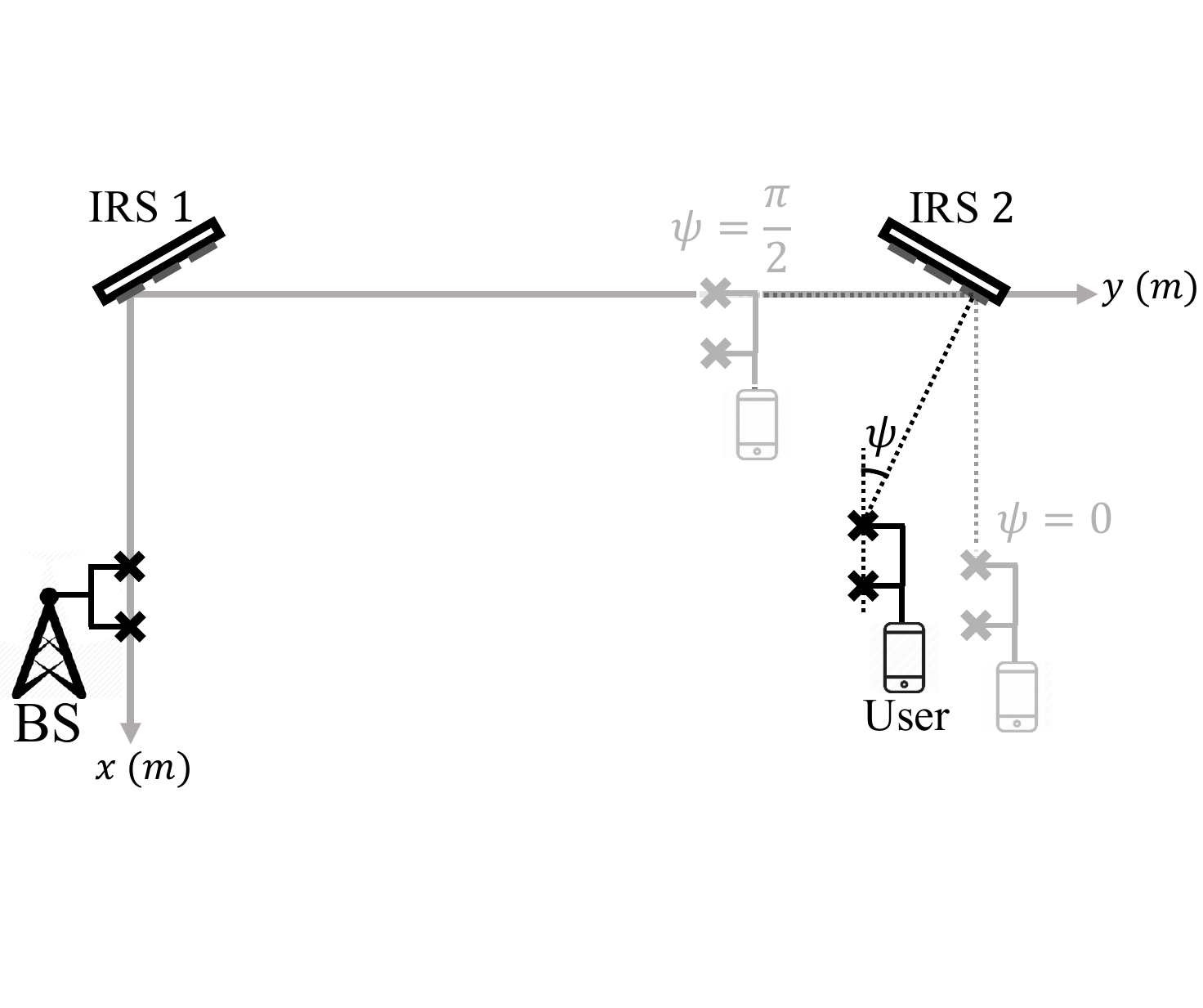}}
\hspace{0.01in}
\subfigure[Achievable rate versus angle $\psi$.]{
\includegraphics[width=3.0in]{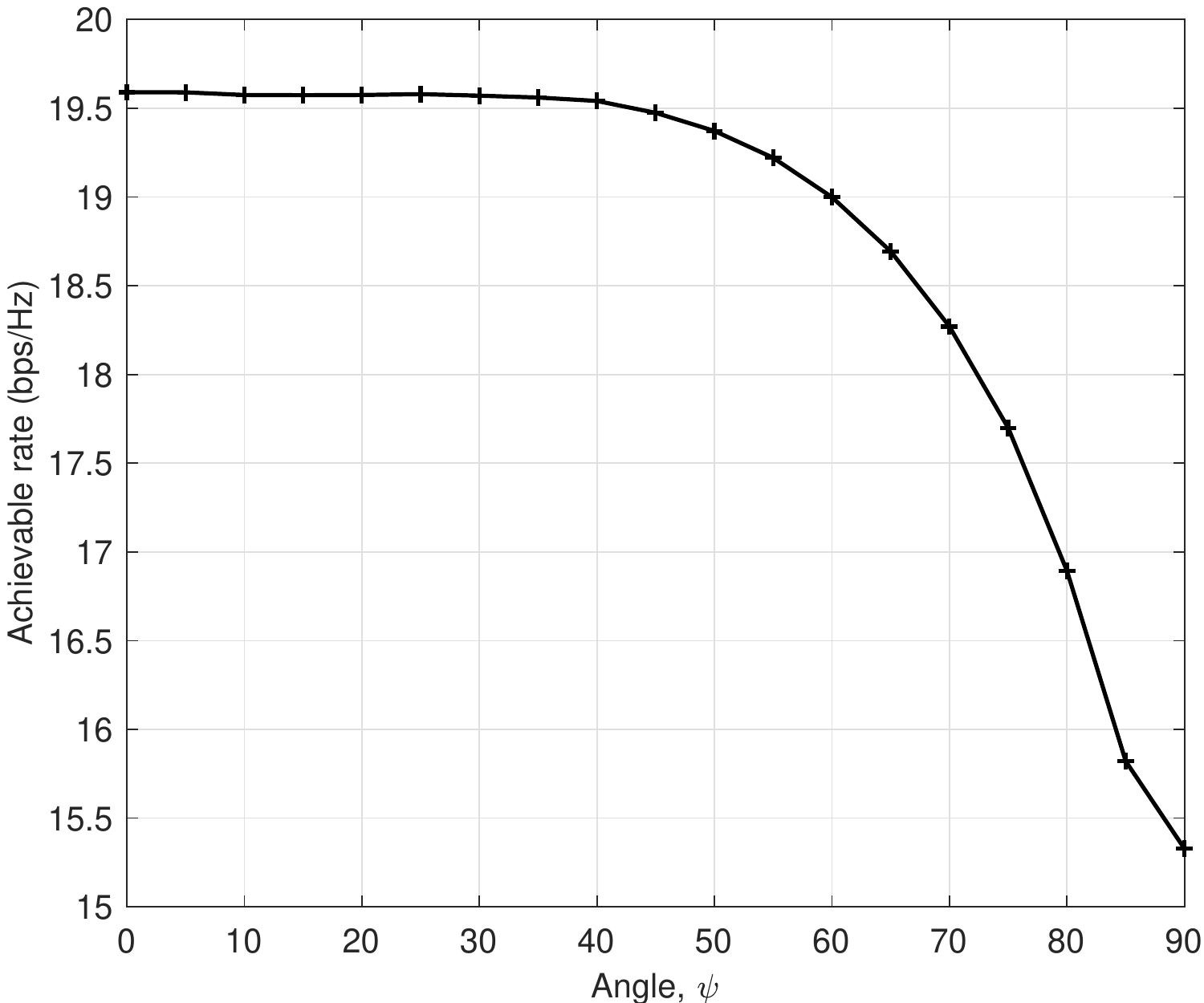}}
\caption{Achievable rate under different user-IRS angles.}
\vspace{-1.8em}
\label{anal}
\end{figure*}

As discussed in Section V, unlike the single-IRS aided MIMO system with only one single-reflection link, the performance of the double-IRS aided MIMO system depends on the angles among the antenna/element arrays at the BS, user, and two IRSs, which determine the superposition effect of the three reflection links characterized by the angle-determined array responses in \eqref{21}. Motivated by this, we aim to evaluate the achievable rate under different user-IRS angles, in order to identify the region that the user can be best served by the double-IRS system. Specifically, as illustrated in Fig.~\ref{anal}(a), we consider the same BS and IRS locations as stated at the beginning of this section, while the user location is set as $\mathbf{u}_r=[\cos(\psi),50-\sin(\psi),0]^T$ with varying $\psi\in[0,\pi/2]$. Note that the distance between IRS $2$ and the user is still $d_{R_2}=1$ m, and $\psi$ determines the angle between the user's ULA and the IRS$2$-user line. We consider the setup with $N_t=N_r=2$, $M=1000$, and a high transmit power $P=20$ dBm.

In Fig.~\ref{anal}(b), we show the achievable rate versus the angle $\psi$. It can be observed that the achievable rate decreases as the angle $\psi$ increases, which can be explained as follows. Note that under the considered setup, the two array responses at the user are given by $\mathbf{r}_{1L}=\big[1,\exp\big(\frac{j2\pi}{\lambda}l_n\cos(\frac{\pi}{2})\big)\big]^T=[1,1]^T$ and $\mathbf{r}_{2L}=\big[1,\exp\big(\frac{j2\pi}{\lambda}l_n\cos(\pi-\psi)\big)\big]^T=[1,\exp(-j\pi\cos(\psi))]^T$, respectively. Recall from Section V that the asymptotic capacity of the double-IRS aided MIMO system in \eqref{45} decreases as the correlation between the array responses at the user, $\rho_{2}(\Theta_r)=\frac{1}{2}\big|\mathbf{r}_{2L}^H\mathbf{r}_{1L}\big|=\frac{1}{2}\big|1+\exp(-j\pi\cos(\psi))\big|$, increases, and $\rho_{2}(\Theta_r)$ is an increasing function of $\psi\in[0,\pi/2]$. Particularly, for the best case of $\psi=0$ as illustrated in Fig. 7 (a), the correlation $\rho_{2}(\Theta_r)$ is minimized to $0$, thus $\mathbf{r}_{1L}$ and $\mathbf{r}_{2L}$ are orthogonal and the achievable rate is maximized (note that now the array responses at the BS $\mathbf{t}_{1R}$ and $\mathbf{t}_{2R}$ are also orthogonal). While for the worst case of $\psi=\frac{\pi}{2}$, the correlation  $\rho_{2}(\Theta_r)$ is maximized to $1$ and the effective MIMO channel $\mathbf{H}$ is of rank one, which leads to the lowest achievable rate since the spatial multiplexing gain cannot be harvested. This suggests that in the high transmit power regime, the two IRSs should be carefully placed such that the array responses at the user are orthogonal, so are the array responses at the BS, for maximizing the MIMO channel capacity.

\section{Conclusion}

In this paper, we investigated a novel double-IRS aided MIMO communication system, where a multi-antenna user is served by a multi-antenna BS through two single-reflection links and one double-reflection link, under the LoS propagation channels. We formulated the capacity maximization problem by jointly optimizing the transmit covariance matrix and the passive beamforming matrices of the two IRSs, which is non-convex and difficult to solve. Nevertheless, by exploiting the unique characteristics of the LoS channels, we proposed a low-complexity algorithm whose complexity is independent with the total number of IRS elements. Then, we analytically showed that the rank of the effective MIMO channel can be increased to two by properly deploying the two IRSs. By considering the case with two antennas at the BS or user, we further analyzed the capacity scaling orders of the double-IRS aided MIMO system with respect to asymptotically large number of IRS elements or transmit power, which are shown to significantly outperform its single-IRS counterpart. Finally, we conducted extensive simulations to evaluate the performance of our proposed algorithm, validate our analytical results, and reveal the advantages of our considered double-IRS system over the conventional single-IRS system.

In future work, it will be interesting to extend the capacity scaling results for the double-IRS aided MIMO system with arbitrary number of transmit/receive antennas, as well as under multi-path/non-LoS channels.   

\section*{Appendices}

\subsection{Proof of Proposition 2}

The MIMO channel capacity for rank-two $\mathbf{H}$ can be written as
\vspace{-0.5em}
\begin{equation}
\vspace{-0.5em}
\begin{aligned}
C&=\log_2\bigg(1+\frac{P_1^\star\delta_1^2}{\sigma^2}+\frac{P_2^\star\delta_2^2}{\sigma^2}+\frac{P_1^\star P_2^\star\delta_1^2\delta_2^2}{\sigma^4}\bigg).
\label{52}
\end{aligned}
\end{equation}For maximizing the capacity scaling order with respect to an asymptotically large $M$, equivalently we are to maximize the highest exponent of $M$ inside \eqref{52}'s logarithm operator. We start with $\frac{P_1^\star P_2^\star\delta_1^2\delta_2^2}{\sigma^4}$, where the expression of $(\delta_1\delta_2)^2$ is given by
\vspace{-0.5em}
\begin{equation}
\vspace{-0.6em}
\begin{aligned}
&(\delta_1\delta_2)^2=\Big|\det\big(\mathbf{H}\mathbf{H}^H\big)\Big|\\
&=\!\Big|a(\mathbf{\Phi}_1)b(\mathbf{\Phi}_2)\Big|^2\bigg|\!\bigg(\!N_t^2e^{-j2\pi(N_t-1)\Theta_{t}}\!-\!\Big(\!\sum_{n_t=1}^{N_t}e^{-j2\pi(n_t-1)\Theta_{t}}\Big)^2\!\bigg)\bigg(\!4e^{-j2\pi\Theta_{r}}\!-\!\Big(\!\sum_{n_r=1}^{2}e^{-j2\pi(n_r-1)\Theta_{r}}\Big)^2\!\bigg)\bigg|\\
&=\Big|a(\mathbf{\Phi}_1)b(\mathbf{\Phi}_2)\Big|^2\bigg|N_t^2e^{-j2\pi(N_t-1)\Theta_{t}}-\Big(\frac{1-e^{-j2\pi N_t\Theta_{t}}}{1-e^{-j2\pi \Theta_{t}}}\Big)^2\bigg|\bigg|4e^{-j2\pi\Theta_{r}}-\Big(\frac{1-e^{-j4\pi\Theta_{r}}}{1-e^{-j2\pi\Theta_{r}}}\Big)^2\bigg|\\
&=\Big|a(\mathbf{\Phi}_1)b(\mathbf{\Phi}_2)\Big|^2\bigg|N_t^2-\Big(\frac{e^{j\pi N_t\Theta_{t}}-e^{-j\pi N_t\Theta_{t}}}{e^{j\pi \Theta_{t}}-e^{-j\pi \Theta_{t}}}\Big)^2\bigg|\bigg|4-\Big(\frac{e^{j2\pi\Theta_{r}}-e^{-j2\pi\Theta_{r}}}{e^{j\pi\Theta_{r}}-e^{-j\pi\Theta_{r}}}\Big)^2\bigg|
\nonumber
\end{aligned}
\end{equation}
\begin{equation}
\vspace{-0.6em}
\begin{aligned}
\!\!\!\!\!\!\!\!\!\!\!\!\!\!\!\!\!\!\!\!\!\!\!\!\!\!\!\!\!\!\!\!\!\!\!\!\!\!\!\!\!\!\!\!\!\!\!\!\!\!\!\!\!\!\!\!\!\!\!\!\!\!\!\!\!\!\!\!\!\!\!\!\!\!\!\!\!\!\!\!\!\!\!\!\!\!\!\!\!\!\!\!\!\!\!\!\!\!\!\!\!\!\!\!=4N_t^2\Big|a(\mathbf{\Phi}_1)b(\mathbf{\Phi}_2)\Big|^2\big(1-\rho_{N_t}(\Theta_{t})^2\big)\big(1-\rho_{2}(\Theta_{r})^2\big).
\label{53}
\end{aligned}
\end{equation}Clearly, by deploying the passive beamforming structure in \eqref{22}, we can maximize $\big|a(\mathbf{\Phi}_1)b(\mathbf{\Phi}_2)\big|^2$ in \eqref{53} as $\big(\frac{\alpha^2 M_1M_2}{d_{R_1}\!d_{T_1}\!d_{R_2}\!d_{T_2}}\big)^2$. By further setting $M_1=M_2=\frac{M}{2}$, $(\delta_1\delta_2)^2$'s highest exponent of $M$ is maximized as $4$; denoting $e_1$ as $\delta_1^2$'s highest exponent of $M$ and $e_2$ as $\delta_2^2$'s highest exponent of $M$, we thus have the following relationship from \eqref{53}
\vspace{-1.0em}
\begin{equation}
\vspace{-0.6em}
\begin{aligned}
e_1+e_2 = 4.
\label{54}
\end{aligned}
\end{equation}

While for $\frac{P_1^\star\delta_1^2}{\sigma^2}$ and $\frac{P_2^\star\delta_2^2}{\sigma^2}$, we have
\vspace{-0.5em}
\begin{equation}
\vspace{-0.8em}
\begin{aligned}
\delta_1^2+\delta_2^2 = \tr\big(\mathbf{H}\mathbf{H}^H\big) = \sum_{n_r\in\mathcal{N}_r,n_t\in\mathcal{N}_t}\big|(\mathbf{H})_{n_r,n_t}\big|^2,
\label{55}
\end{aligned}
\end{equation}where the entries of $\mathbf{H}$ are given by
\vspace{-0.5em}
\begin{equation}
\vspace{-0.5em}
\begin{aligned}
(\mathbf{H})_{n_r,n_t}\!\!=\!\!\ a(\mathbf{\Phi}_1)&e^{\frac{-j2\pi l_n}{\lambda}\!\big(\!(\!n_{r}-1\!)\!\cos(\omega_{R_1,v_r}\!)-(\!n_{t}-1\!)\cos(\omega_{T_1,v_t}\!)\!\big)}\!+\!b(\mathbf{\Phi}_2)e^{\frac{-j2\pi l_n}{\lambda}\!\big(\!(\!n_{r}-1\!)\!\cos(\omega_{R_2,v_r}\!)-(\!n_{t}-1\!)\!\cos(\omega_{T_2,v_t}\!)\!\big)}\nonumber
\end{aligned}
\end{equation}
\begin{equation}
\vspace{-0.5em}
\begin{aligned}
&+\check{c}(\mathbf{\Phi}_1,\mathbf{\Phi}_2)e^{\frac{-j2\pi l_n}{\lambda}\big((n_{r}-1)\cos(\omega_{R_2,v_r})-(n_{t}-1)\cos(\omega_{T_1,v_t})\big)},\ n_r\in\mathcal{N}_r,n_t\in\mathcal{N}_t.
\label{56}
\end{aligned}
\end{equation}Hence, by adopting the same passive beamforming structure in \eqref{22} together with $M_1=M_2=\frac{M}{2}$, the highest exponent of $M$ of each $\big|(\mathbf{H})_{n_r,n_t}\big|^2$ is maximized as $4$, in turn that of the RHS of \eqref{55} is also maximized as $4$, and we have the following relationship from \eqref{55}:
\vspace{-0.7em}
\begin{equation}
\vspace{-0.5em}
\begin{aligned}
\max(e_1,e_2)\geq4.
\label{57}
\end{aligned}
\end{equation}Next, we show that $\max(e_1,e_2)>4$ cannot hold by contradiction. Specifically, $e_1=e_2$ needs to hold for $\max(e_1,e_2)>4$, since the terms in $\delta_1^2$ with $M$'s exponent higher than $4$ must have the same amplitudes but opposite signs with those in $\delta_2^2$, so that they can be cancelled out on the LHS of \eqref{55} and the highest exponent of $M$ on the RHS of \eqref{55} is still $4$. In this case, $e_1+e_2=2e_1>4$, which contradicts with \eqref{54}. Therefore, we must have $\max(e_1,e_2)=4$. Particularly, recall that $\delta_1$ is larger than $\delta_2$, we thus have $e_1=4$ and $e_2=0$, i.e., $\delta_1^2$ scales with $\mathcal{O}(M^4)$ and $\delta_2^2$ is a constant for asymptotically large $M$.

Note that given any feasible water-filling power allocation $\{P_1^\star,P_2^\star\}$, the strongest eigenchannel will always be allocated with a strictly positive transmit power, i.e., $P_1^\star>0$, and we have
\begin{itemize}
\item{If $P_1^\star>0$ and $P_2^\star>0$, then
\vspace{-0.5em}
\begin{equation}
\vspace{-0.5em}
\begin{aligned}
C&=\log_2\bigg(1+\frac{P_1^\star\delta_1^2}{\sigma^2}+\frac{P_2^\star\delta_2^2}{\sigma^2}+\frac{P_1^\star P_2^\star\delta_1^2\delta_2^2}{\sigma^4}\bigg),
\label{58}
\end{aligned}
\end{equation}and the highest exponent of $M$ inside \eqref{58}'s logarithm operator is $4$.
}
\item{If $P_1^\star=P$ and $P_2^\star=0$, then
\vspace{-0.5em}
\begin{equation}
\vspace{-0.5em}
\begin{aligned}
C&=\log_2\bigg(1+\frac{P\delta_1^2}{\sigma^2}\bigg),
\label{59}
\end{aligned}
\end{equation}and the highest exponent of $M$ inside \eqref{59}'s logarithm operator is also $4$.
}
\end{itemize}
To summarize, the highest capacity scaling order with respect to an asymptotically large $M$ is $4$, i.e., $\underset{M\rightarrow\infty}{\lim}\frac{C}{\log_2(M)}=4$, by deploying the passive beamforming structure in \eqref{22}. Proposition 2 is thus proved. 

\subsection{Proof of Proposition 4}

The MIMO channel capacity for rank-one $\mathbf{H}$ can be written as
\vspace{-0.5em}
\begin{equation}
\vspace{-0.5em}
\begin{aligned}
C&=\log_2\bigg(1+\frac{P\delta_1^2}{\sigma^2}\bigg).
\label{60}
\end{aligned}
\end{equation}While the expression of $\delta_1^2$ for the rank-one $\mathbf{H}$ in \eqref{47} is given by
\vspace{-0.5em}
\begin{equation}
\vspace{-0.5em}
\begin{aligned}
&\delta_1^2 = \tr\big(\mathbf{H}\mathbf{H}^H\big) = \sum_{n_r\in\mathcal{N}_r,n_t\in\mathcal{N}_t}\big|(\mathbf{H})_{n_r,n_t}\big|^2\\
& =N_t\sum_{n_r=1}^{2}\Big|a(\mathbf{\Phi}_1)e^{\frac{-j2\pi}{\lambda}(n_{r}-1)l_{n}\cos(\omega_{R_1,v_r})}+\big(b(\mathbf{\Phi}_2)\!+\!\check{c}(\mathbf{\Phi}_1,\mathbf{\Phi}_2)\big)e^{\frac{-j2\pi}{\lambda}(n_{r}-1)l_{n}\cos(\omega_{R_2,v_r})}\Big|^2\\
&\overset{\text{(a)}}{=}N_t\sum_{n_r=1}^{2}\bigg|\big|a(\mathbf{\Phi}_1)\big|e^{j\big(\frac{-2\pi}{\lambda}(n_{r}-1)l_{n}\cos(\omega_{R_1,v_r})+\varphi_a\big)}+\big|b(\mathbf{\Phi}_2)\!+\!\check{c}(\mathbf{\Phi}_1,\mathbf{\Phi}_2)\big|e^{j\big(\frac{-2\pi}{\lambda}(n_{r}-1)l_{n}\cos(\omega_{R_2,v_r})+\varphi_{b+c}\big)}\bigg|^2\\
&=N_t\sum_{n_r=1}^{2}\Bigg(\bigg(\big|a(\mathbf{\Phi}_1)\big|\cos\Big(\frac{-2\pi}{\lambda}(n_{r}-1)l_{n}\cos(\omega_{R_1,v_r})+\varphi_a\Big)\\
&\ \ \ \ \ \ \ \ \ \ \ \ \ \ \ \ \ \ \ \ +\big|b(\mathbf{\Phi}_2)\!+\!\check{c}(\mathbf{\Phi}_1,\mathbf{\Phi}_2)\big|\cos\Big(\frac{-2\pi}{\lambda}(n_{r}-1)l_{n}\cos(\omega_{R_2,v_r})+\varphi_{b+c}\Big)\bigg)^2\\
&\ \ \ \ \ \ \ \ \ \ \ \ +\bigg(\big|a(\mathbf{\Phi}_1)\big|\sin\Big(\frac{-2\pi}{\lambda}(n_{r}-1)l_{n}\cos(\omega_{R_1,v_r})+\varphi_a\Big)\\
&\ \ \ \ \ \ \ \ \ \ \ \ \ \ \ \ \ \ \ \ +\big|b(\mathbf{\Phi}_2)\!+\!\check{c}(\mathbf{\Phi}_1,\mathbf{\Phi}_2)\big|\sin\Big(\frac{-2\pi}{\lambda}(n_{r}-1)l_{n}\cos(\omega_{R_2,v_r})+\varphi_{b+c}\Big)\bigg)^2\Bigg)\nonumber
\end{aligned}
\end{equation}
\vspace{-0.5em}
\begin{equation}
\vspace{-0.5em}
\begin{aligned}
\!\!\!\!\!\!\!\!\!\!\!\!\!\!\!\!\!\!\!\!\!\!\!=2N_t\Big(\big|a(\mathbf{\Phi}_1)\big|^2\!+\!\big|b(\mathbf{\Phi}_2)\!+\!\check{c}(\mathbf{\Phi}_1,\mathbf{\Phi}_2)\big|^2\Big)+&2N_t\big|a(\mathbf{\Phi}_1)\big|\big|b(\mathbf{\Phi}_2)\!+\!\check{c}(\mathbf{\Phi}_1,\mathbf{\Phi}_2)\big|\\
&\times\sum_{n_r=1}^{2}\cos\Big(2\pi(n_{r}-1)\Theta_r+(\varphi_{b+c}-\varphi_{a})\Big)\nonumber
\end{aligned}
\end{equation}
\vspace{-0.5em}
\begin{equation}
\vspace{-0.5em}
\begin{aligned}
\!\!\!\!\!\!\!\!\!\!\!\!\!\!\!\!\!\!\!\!\!\!\!\!\!\!\!\!\!\!\!\!\!\!\!\!\!=2N_t\Big(\big|a(\mathbf{\Phi}_1)\big|^2\!+\!\big|b(\mathbf{\Phi}_2)\!+\!\check{c}(\mathbf{\Phi}_1,\mathbf{\Phi}_2)\big|^2\Big)+&4N_t\big|a(\mathbf{\Phi}_1)\big|\big|b(\mathbf{\Phi}_2)\!+\!\check{c}(\mathbf{\Phi}_1,\mathbf{\Phi}_2)\big|\\
&\times\rho_{2}(\Theta_r)\cos\Big(\pi\Theta_{r}+\big(\varphi_{b+c}-\varphi_{a}\big)\Big),
\label{61}
\end{aligned}
\end{equation}where $\overset{\text{(a)}}{=}$ comes from denoting $\varphi_a=\arg\big(a(\mathbf{\Phi}_1)\big)$ and $\varphi_{b+c}=\arg\big(b(\mathbf{\Phi}_2)\!+\!\check{c}(\mathbf{\Phi}_1,\mathbf{\Phi}_2)\big)$.

To maximize $\delta_1^2$ in \eqref{61}, we need to maximize $|a(\mathbf{\Phi}_1)|$, $|b(\mathbf{\Phi}_2)|$, and $|\check{c}(\mathbf{\Phi}_1,\mathbf{\Phi}_2)|$ by employing the passive beamforming structure in \eqref{22}, which gives
\vspace{-0.5em}
\begin{equation}
\vspace{-0.5em}
\begin{aligned}
\delta_1^2&=2N_t\bigg(\bigg|\frac{\alpha M_1\beta_a\gamma_1}{d_{R_1}\!d_{T_1}}\bigg|^2\!+\!\bigg|\frac{\alpha M_2\beta_b\gamma_2}{d_{R_2}\!d_{T_2}}\!+\!\frac{\alpha^{3/2} M_1M_2\beta_c\gamma_1\gamma_2}{d_{R_2}\!d_{S}d_{T_1}}\bigg|^2\bigg)+4N_t\bigg|\frac{\alpha M_1\beta_a\gamma_1}{d_{R_1}\!d_{T_1}}\bigg|\\
&\ \ \ \ \ \ \ \ \ \ \ \ \ \ \ \ \ \ \ \ \ \times\bigg|\frac{\alpha M_2\beta_b\gamma_2}{d_{R_2}\!d_{T_2}}\!+\!\frac{\alpha^{3/2} M_1M_2\beta_c\gamma_1\gamma_2}{d_{R_2}\!d_{S}d_{T_1}}\bigg|\rho_{2}(\Theta_r)\cos\Big(\pi\Theta_{r}+\big(\varphi_{b+c}-\varphi_{a}\big)\Big).
\label{62}
\end{aligned}
\end{equation}Clearly, we should optimally set the common phase shift of IRS $1$ as $\gamma_1^\star=\frac{\beta_b}{\beta_c}$ for maximizing $\big|\frac{\alpha M_2\beta_b\gamma_2}{d_{R_2}\!d_{T_2}}\!+\!\frac{\alpha^{3/2} M_1M_2\beta_c\gamma_1\gamma_2}{d_{R_2}\!d_{S}d_{T_1}}\big|$, which yields $\varphi_a=\frac{\beta_a\beta_b}{\beta_c}$ and $\varphi_{b+c}=\beta_b\gamma_2$. We should further optimally set the common phase shift of IRS $2$ as $\gamma_2^\star=\frac{1}{\beta_b}\big(\frac{\beta_a\beta_b}{\beta_c}-\pi\Theta_r\big)$ for maximizing the cosine term in \eqref{62}. Therefore, the MIMO channel capacity in \eqref{60} is maximized as
\vspace{-0.5em}
\begin{equation}
\vspace{-0.5em}
\begin{aligned}
C\!\!=\!\log_2\!\!\Bigg(\!1\!\!+\!\frac{2N_tP}{\sigma^2}\!\bigg(\!\!\Big(\frac{\alpha M_1}{d_{R_1}\!d_{T_1}}\!\Big)^2\!\!\!+\!\Big(\frac{\alpha M_2}{d_{R_2}\!d_{T_2}}\!+\!\frac{\alpha^{3/2} M_1\!M_2}{d_{R_2}\!d_{S}d_{T_1}}\!\Big)^2\!\!\!+\!2\Big(\frac{\alpha M_1}{d_{R_1}\!d_{T_1}}\!\Big)\Big(\frac{\alpha M_2}{d_{R_2}\!d_{T_2}}\!+\!\frac{\alpha^{3/2} M_1\!M_2}{d_{R_2}\!d_{S}d_{T_1}}\!\Big)\rho_{2}(\Theta_r)\!\bigg)\!\!\Bigg).
\label{63}
\end{aligned}
\end{equation}This thus completes the proof of Proposition 4.

\end{spacing}

\end{document}